\documentclass{ws-ijmpa-mod}
\usepackage[super,compress]{cite}
\usepackage{graphicx,hyperref}
\usepackage{url}
\usepackage[caption=false]{subfig}
\DeclareGraphicsExtensions{.pdf,.png}

\pdfoutput=1 

\def\t#1{\tilde{ #1}}
\def\tev{\textrm{TeV}}
\def\gev{\textrm{GeV}}

\def\ifb{\ensuremath{\textrm{fb}^{-1}}}

\def\W{\ensuremath{W}}
\def\Z{\ensuremath{Z}}
\def\to{\ensuremath{\rightarrow}}

\def\X{\ensuremath{\tilde\chi_1^0}}
\def\x{\ensuremath{\chi}}
\def\mx{\ensuremath{m_{\chi}}}
\def\R{\emph{R}}

\def\pt{\ensuremath{p_{\rm T}}}
\def\et{\ensuremath{E_{\rm T}}}
\def\met{\ensuremath{E_{\rm T}^{\rm miss}}}
\def\mef{\ensuremath{M_{\rm eff}}}

\def\mt{\ensuremath{M_{\rm T}}}

\begin{document}
\markboth{V.~A.~Mitsou}
{Shedding Light on Dark Matter at Colliders}

%%%%%%%%%%%%%%%%%%%%% Publisher's Area please ignore %%%%%%%%%%%%%%%
%
\catchline{}{}{}{}{}
%
%%%%%%%%%%%%%%%%%%%%%%%%%%%%%%%%%%%%%%%%%%%%%%%%%%%%%%%%%%%%%%%%%%%%

\title{SHEDDING LIGHT ON DARK MATTER AT COLLIDERS}

\author{VASILIKI A.\ MITSOU}

\address{Instituto de F\'isica Corpuscular (IFIC), CSIC -- Universitat de Val\`encia, \\
Parc Cient\'ific de la U.V., C/ Catedr\'atico Jos\'e Beltr\'an 2,
E-46980 Paterna (Valencia), Spain \\
vasiliki.mitsou@ific.uv.es}

\maketitle

\begin{history}
\received{Day Month Year}
\revised{Day Month Year}
\end{history}

\begin{abstract}
Dark matter remains one of the most puzzling mysteries in Fundamental Physics of our times. Experiments at high-energy physics colliders are expected to shed light to its nature and determine its properties. This review focuses on recent searches for dark-matter signatures at the Large Hadron Collider, also discussing related prospects in future $e^+e^-$ colliders. 

\end{abstract}

\ccode{PACS numbers: 95.35.+d, 12.60.Jv, 14.70.Kv}
% Dark matter, Supersymmetric models, Gravitons 

\tableofcontents

%%%%%%%%%%%%%%%%%%%%%%%%%%%%%%%%%%%%%%%%%%%%%%%%%%%%%%%%%%%%%%%%%%%%%%%%%%%%%%%%%%%%%%%%%%%%%%%%%%%%%%%%%%%%
%%%%%%%%%%%%%%%%%%%%%%%%%%%%%%%%%%%%%%%%%%%%%%%%%%%%%%%%%%%%%%%%%%%%%%%%%%%%%%%%%%%%%%%%%%%%%%%%%%%%%%%%%%%%
\section{Introduction}\label{sc:intro}
%%%%%%%%%%%%%%%%%%%%%%%%%%%%%%%%%%%%%%%%%%%%%%%%%%%%%%%%%%%%%%%%%%%%%%%%%%%%%%%%%%%%%%%%%%%%%%%%%%%%%%%%%%%%
%%%%%%%%%%%%%%%%%%%%%%%%%%%%%%%%%%%%%%%%%%%%%%%%%%%%%%%%%%%%%%%%%%%%%%%%%%%%%%%%%%%%%%%%%%%%%%%%%%%%%%%%%%%%

Unveiling the nature of dark matter (DM)~\cite{dm-review} is a quest in both Astroparticle and Particle Physics. Among the list of well-motivated candidates, the most popular particles are \emph{cold} and weakly interacting, and typically predict missing-energy signals at particle colliders. Supersymmetry~\cite{susy-dm1,susy-dm2,susy-dm3,susy-dm4,susy-dm5,susy-dm6,susy-dm7,susy-dm8,susy-dm9} and models with extra dimensions~\cite{ued-dm} are theoretical scenarios that inherently provide such a dark matter candidate. High-energy colliders, such as the CERN Large Hadron Collider~\cite{lhc}, are ideal machines for producing and eventually detecting DM. Experiments in upcoming colliders, such as the ILC~\cite{ilc} and CLIC~\cite{clic}, are expected to further constraint such models, should they are materialized in Nature, and subsequently make a key step in understanding dark matter.  

In parallel, the exploration of dark matter is being pursued through other types of detection methods: direct detection in low-background underground experiments~\cite{direct1,direct2} and indirect detection of neutrinos, $\gamma$-rays and antimatter with terrestrial and space-borne detectors~\cite{indirect}. A recent review on the status and results of these instruments is given in Ref.~\citen{lisboa}.

The structure of this paper is as follows. Section~\ref{sc:intro} provides a brief introduction to the properties of dark matter as defined by the current cosmological data and its implications for physics in colliders. Section~\ref{sc:lhc} highlights the features of the LHC experiments that play a central role in exploring DM. In Section~\ref{sc:monox}, the strategy, methods, and results of the LHC experiments as far as model-independent DM-production is concerned are discussed. In Sections~\ref{sc:susy} and~\ref{sc:ed}, the latest results in searches for supersymmetry and for extra dimensions, respectively, are presented. The prospects for exploring dark matter and possibly measuring its properties at future colliders are given in Section~\ref{sc:ilc}. The paper concludes with a summary and an outlook in Section~\ref{sc:sums}.

%%%%%%%%%%%%%%%%%%%%%%%%%%%%%%%%%%%%%%%%%%%%%%%%%%%%%%%%%%%%%%%%%%%%%%%%%%%%%%%%%%%%%%%%%%%%%%%%%%%%%%%%%%%%
\subsection{Dark matter evidence}\label{sc:dm}
%%%%%%%%%%%%%%%%%%%%%%%%%%%%%%%%%%%%%%%%%%%%%%%%%%%%%%%%%%%%%%%%%%%%%%%%%%%%%%%%%%%%%%%%%%%%%%%%%%%%%%%%%%%%

The nature of the dark sector of the Universe constitutes one of the major mysteries of fundamental physics. According to the latest observations by the Planck mission team~\cite{planck2}, most of our Universe energy budget consists of unknown entities: $\sim\!26.8\%$ is dark matter and $\sim\!68.3\%$ is dark energy, a form of ground-state energy. Dark energy is believed to be responsible for the current-era acceleration of the Universe. Dark matter, on the other hand, is matter inferred to exist from gravitational effects on visible matter, being undetectable by emitted or scattered electromagnetic radiation. A possible explanation other than the introduction of one or more yet-unknown particles is to ascribe the observed effects to modified Newtonian dynamics~\cite{mond1,mond2}. 

\begin{figure}[ht]
\begin{minipage}[b]{0.33\textwidth}
\centerline{\includegraphics[width=\textwidth]{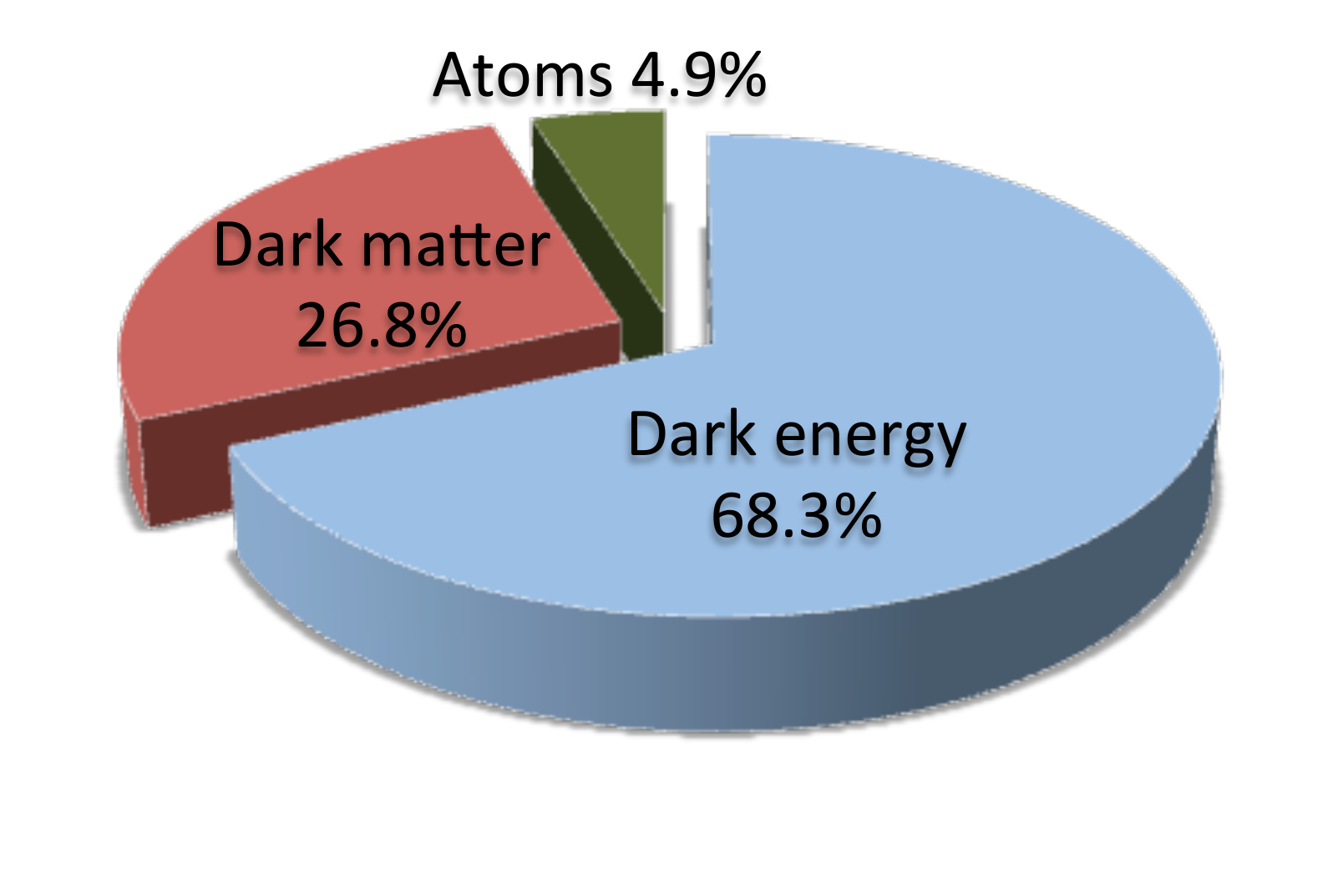}}
\caption{\label{fg:budget}The energy budget of the Universe according to recent cosmological evidence~\cite{planck2} and assuming the $\Lambda$CDM model~\cite{lcdm}.}
\end{minipage}\hfill %\hspace{0.04\textwidth}%
\begin{minipage}[b]{0.62\textwidth}
\centerline{\includegraphics[width=\textwidth]{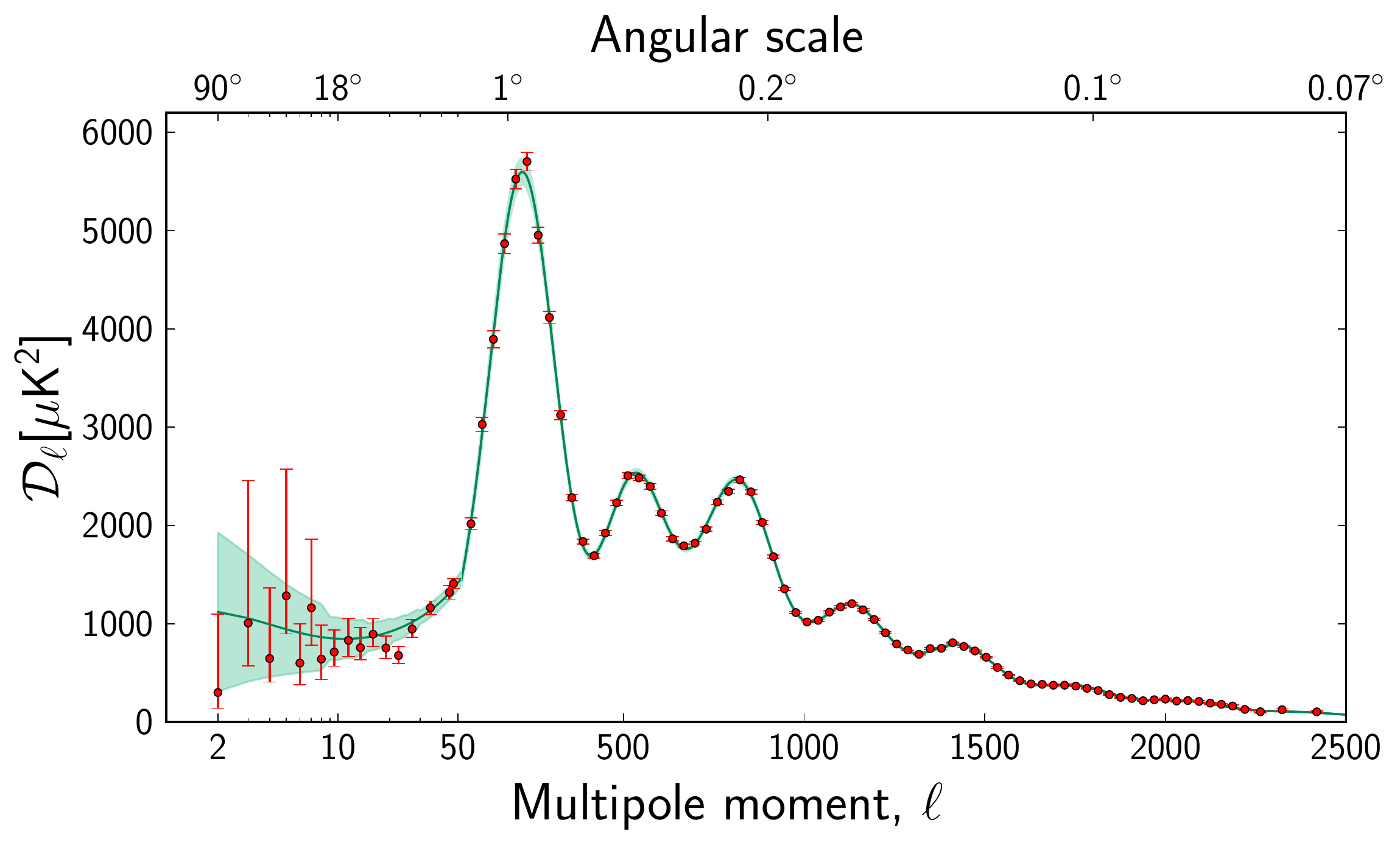}}
\caption{\label{fg:cmb}Temperature power spectrum from Planck. The densities of baryoninc and dark matter are measured from the relative heights of the acoustic peaks. The third acoustic peak is sensitive to the dark matter density. From Ref.~\citen{planck1}.}
\end{minipage} 
\end{figure}

The energy budget of the Cosmos (Fig.~\ref{fg:budget}) has been obtained by combining a variety of astrophysical data, such as type-Ia supernovae~\cite{snIa1,snIa2,snIa3}, cosmic microwave background (CMB)~\cite{wmap,planck1}, baryonic acoustic oscillations~\cite{bao1,bao2} and weak-lensing data~\cite{lensing}. The most precise measurement comes from anisotropies of the cosmic microwave background, as reflected in the its power spectrum, shown in Fig.~\ref{fg:cmb}.

Evidence from the formation of large-scale structure (galaxies and their clusters) strongly favor cosmologies where non-baryonic DM is entirely composed of cold dark matter (CDM), i.e.\ non-relativistic particles. CDM particles, in turn, may be axions~\cite{axion}, superheavy non-thermal relics (wimpzillas, cryptons)~\cite{shdm} or weakly interacting massive particles (WIMPs). The latter class of DM candidates arises naturally in models which attempt to explain the origin of electroweak symmetry breaking and this is precisely where the connection between Cosmology and Particle Physics lies. Furthermore, the typical (weak-scale) cross sections characterizing these models are of the same order of magnitude as the WIMP annihilation cross section, thus establishing the so-called \emph{WIMP miracle}. A review on the interplay between (string-inspired) Cosmology and the LHC with an emphasis on the dark sector is given in Ref.~\citen{mavromatos} and in references therein.

%%%%%%%%%%%%%%%%%%%%%%%%%%%%%%%%%%%%%%%%%%%%%%%%%%%%%%%%%%%%%%%%%%%%%%%%%%%%%%%%%%%%%%%%%%%%%%%%%%%%%%%%%%%%
\subsection{Connection between WIMPs and colliders}\label{sc:colliders}
%%%%%%%%%%%%%%%%%%%%%%%%%%%%%%%%%%%%%%%%%%%%%%%%%%%%%%%%%%%%%%%%%%%%%%%%%%%%%%%%%%%%%%%%%%%%%%%%%%%%%%%%%%%%

WIMP dark matter candidates include the lightest neutralino in models with weak-scale supersymmetry~\cite{susy-dm1,susy-dm2,susy-dm3,susy-dm4,susy-dm5,susy-dm6,susy-dm7,susy-dm8,susy-dm9}, Kaluza-Klein photons arise in scenarios with universal extra dimensions (UED)~\cite{ued-dm}, while lightest $T$-odd particles are predicted in Little Higgs models~\cite{little} with a conserved $T$-parity. The common denominator in these theories is that they all predict the existence of an electrically neutral, colorless and \emph{stable} particle, whose decay is prevented by a kind of symmetry: \R-parity, connected to baryon and lepton number conservation in SUSY models; KK-parity, the four-dimensional remnant of momentum conservation in extra dimension scenarios; and a $Z_2$ discrete symmetry called $T$-parity in Little Higgs models. 

Weakly interacting massive particles do not interact neither electromagnetically nor strongly with matter and thus, once produced, they traverse the various detectors layers without leaving a trace, just like neutrinos. However by exploiting the hermeticity of the experiments, we can get a hint of the WIMP presence through the balance of the energy/momentum measured in the various detector components, the so-called \emph{missing energy}. In hadron colliders, in particular, since the longitudinal momenta of the colliding partons are unknown, only the \emph{transverse missing energy}, \met, can be reliably used to `detect' DM particles. In this paper, we focus on generic DM searches, on supersymmetric signatures, and on possible signals from extra dimensions, all based on \met, performed by the two main LHC experiments, ATLAS~\cite{atlas-det} and CMS~\cite{cms-det}. 

%%%%%%%%%%%%%%%%%%%%%%%%%%%%%%%%%%%%%%%%%%%%%%%%%%%%%%%%%%%%%%%%%%%%%%%%%%%%%%%%%%%%%%%%%%%%%%%%%%%%%%%%%%%%
%%%%%%%%%%%%%%%%%%%%%%%%%%%%%%%%%%%%%%%%%%%%%%%%%%%%%%%%%%%%%%%%%%%%%%%%%%%%%%%%%%%%%%%%%%%%%%%%%%%%%%%%%%%%
\section{The ATLAS and CMS Experiments at the LHC}\label{sc:lhc}
%%%%%%%%%%%%%%%%%%%%%%%%%%%%%%%%%%%%%%%%%%%%%%%%%%%%%%%%%%%%%%%%%%%%%%%%%%%%%%%%%%%%%%%%%%%%%%%%%%%%%%%%%%%%
%%%%%%%%%%%%%%%%%%%%%%%%%%%%%%%%%%%%%%%%%%%%%%%%%%%%%%%%%%%%%%%%%%%%%%%%%%%%%%%%%%%%%%%%%%%%%%%%%%%%%%%%%%%%

The Large Hadron Collider (LHC)~\cite{lhc}, situated at CERN, the European Laboratory for Particle Physics, outside Geneva, Switzerland, started its physics program in 2010 colliding two counter-rotating beams of protons or heavy ions. Before the scheduled 2013--2015 long shutdown, the LHC succeeded in delivering $\sim5~\ifb$ of integrated luminosity at center-of-mass energy of $7~\tev$ during 2010--2011 and another $\sim23~\ifb$ at $\sqrt{s}=8~\tev$ in 2012. The LHC has already extended considerably the reach of its predecessor hadron machine, the Fermilab Tevatron, both in terms of instantaneous luminosity and energy, despite the fact that it has not arrived yet to its design capabilities. 

The two general-purpose experiments, ATLAS (A Toroidal LHC ApparatuS)~\cite{atlas-det} and CMS (Compact Muon Solenoid)~\cite{cms-det}, have been constructed and operate with the aim of exploring a wide range of possible signals of New Physics that LHC renders accessible, on one hand, and performing precision measurements of Standard Model (SM)~\cite{sm1,sm2,sm3} parameters, on the other. Two other large experiments, namely LHCb~\cite{lhcb-det} and ALICE~\cite{alice-det}, are dedicated to $B$-physics and heavy ions, respectively, also run at the LHC. It is worth mentioning that the MoEDAL~\cite{moedal} experiment is specifically designed to explore high-ionization signatures that may also arise in some dark matter theoretical scenarios~\cite{moedal-review}. 

The ATLAS and CMS detectors are designed to overcome difficult experimental challenges: high radiation levels, large interaction rate and extremely small production cross sections of New Physics signals with respect to the well-known SM processes. To this end, both experiments feature separate subsystems to measure charged particle momentum, energy deposited by electromagnetic showers from photons and electrons, energy from hadronic showers of strongly-interacting particles and muon-track momentum. Complete descriptions of the CMS and ATLAS detectors are available in Refs.~\citen{cms-det} and~\citen{atlas-det}, respectively.

The most remarkable highlight of ATLAS and CMS operation so far is undoubtedly the discovery~\cite{higgs-disc1,higgs-disc2} of a new particle that so far seems to have all the features pinpointing to a SM(-like) Higgs boson~\cite{higgs-prop1,higgs-prop2,higgs-prop3}. The observation of this new boson has strong impact not only on our understanding of the fundamental interactions of Nature, as encoded in the SM, but on the proposed theoretical scenarios of Physics beyond the SM (BSM), as we shall see in the following.  

%%%%%%%%%%%%%%%%%%%%%%%%%%%%%%%%%%%%%%%%%%%%%%%%%%%%%%%%%%%%%%%%%%%%%%%%%%%%%%%%%%%%%%%%%%%%%%%%%%%%%%%%%%%%
%%%%%%%%%%%%%%%%%%%%%%%%%%%%%%%%%%%%%%%%%%%%%%%%%%%%%%%%%%%%%%%%%%%%%%%%%%%%%%%%%%%%%%%%%%%%%%%%%%%%%%%%%%%%
\section{Model-Independent DM Production at the LHC}\label{sc:monox}
%%%%%%%%%%%%%%%%%%%%%%%%%%%%%%%%%%%%%%%%%%%%%%%%%%%%%%%%%%%%%%%%%%%%%%%%%%%%%%%%%%%%%%%%%%%%%%%%%%%%%%%%%%%%
%%%%%%%%%%%%%%%%%%%%%%%%%%%%%%%%%%%%%%%%%%%%%%%%%%%%%%%%%%%%%%%%%%%%%%%%%%%%%%%%%%%%%%%%%%%%%%%%%%%%%%%%%%%%

Collider searches for dark matter are highly complementary to direct~\cite{susy-dm1,susy-dm2,susy-dm3,susy-dm4,susy-dm5,susy-dm6,susy-dm7,susy-dm8,susy-dm9,col-direct1,col-direct2,col-di-in1,col-di-in2,col-di-in3} and indirect~\cite{susy-dm1,susy-dm2,susy-dm3,susy-dm4,susy-dm5,susy-dm6,susy-dm7,susy-dm8,susy-dm9,col-di-in1,col-di-in2,col-di-in3,col-indirect} DM detection methods. The main advantage of collider searches is that they do not suffer from astrophysical uncertainties and that there is no lower limit to the DM masses to which they are sensitive.  

The leading generic diagrams responsible for DM production~\cite{tevatron1,maverick,tevatron2} at hadron colliders, as shown in Fig.~\ref{fg:mono-isr}, involve the pair-production of WIMPs plus the initial- or final-state radiation (ISR/FSR) of a gluon, photon or a weak gauge boson $Z,\,W$. The ISR/FSR particle is necessary to balance the two WIMPs' momentum, so that they are not produced back-to-back resulting in negligible \met. Therefore the search is based on selecting events high-\met\ events, due to the WIMPs, and a single jet, photon or boson candidate. A single-jet event from the CMS experiment is visible in Fig.~\ref{fg:cms-monojet-event}. 

\begin{figure}[htb]
  \centering
  \subfloat[$q\bar{q}\to\x\x+g$]{\label{fg:mono-jet}\includegraphics[width=0.27\textwidth]{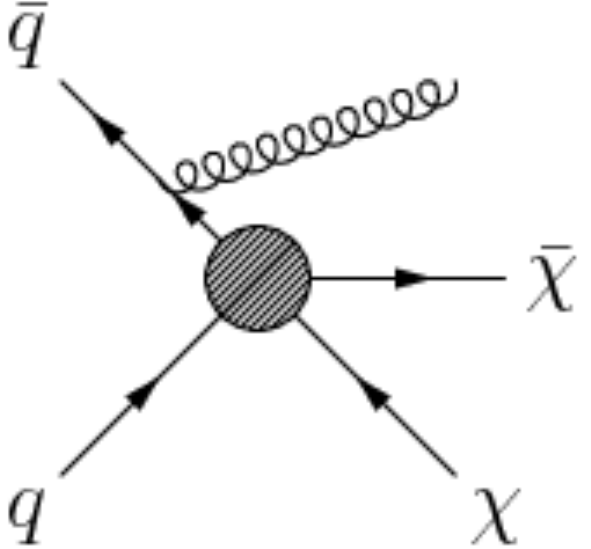}}\qquad
  \subfloat[$q\bar{q}\to\x\x+\gamma, Z, W$]{\label{fg:mono-v}\includegraphics[width=0.31\textwidth]{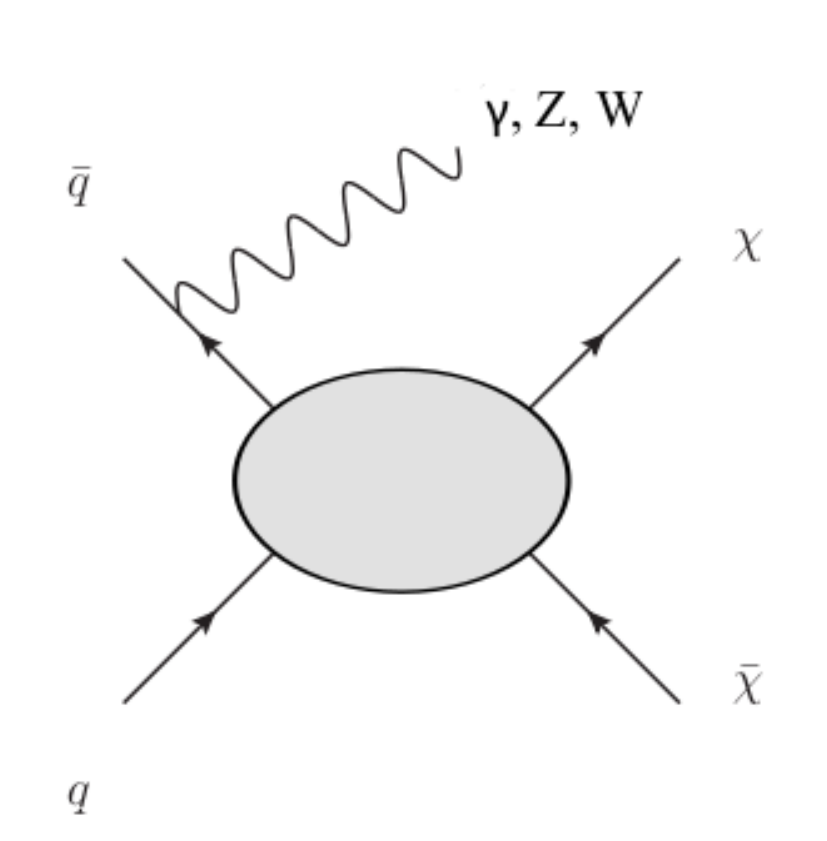}}\\
  \caption{WIMP production at hadron colliders in association with (a) a jet or (b) a photon or a \Z\ or \W\ boson.} \label{fg:mono-isr}
\end{figure}

\begin{figure}[htb]
\centerline{\includegraphics[width=0.7\textwidth]{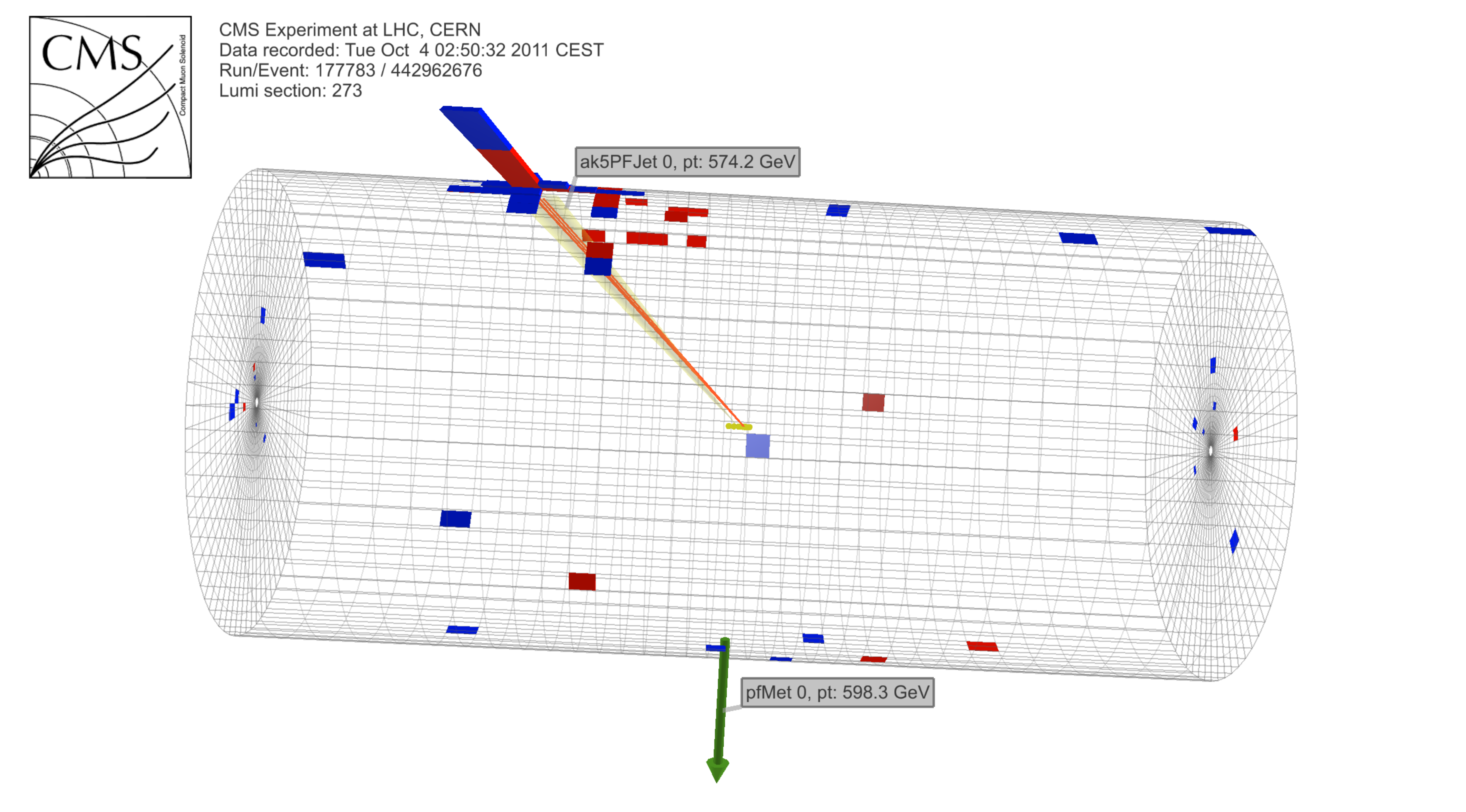}}
\caption{The cylindrical view of a monojet candidate event ($p_{\rm T}^{\rm jet}= 574.2~\gev$, $\met = 598.3~\gev$) from the CMS experiment~\cite{cms-monojet}.  \label{fg:cms-monojet-event}}
\end{figure}

The `blobs' in the diagrams of Fig.~\ref{fg:mono-isr} represent a largely model-independent effective-field-theory framework, in which the interactions between a DM Dirac fermion \x\ and SM fermions $f$ are described by contact operators of the form
\begin{align}
  \mathcal{O}_V &= \frac{(\bar\chi\gamma_\mu\chi)(\bar f \gamma^\mu f)}{M_*^2} \,,
    & \text{($s$-channel, vector)} \label{eq:1} \\
  \mathcal{O}_S &= \frac{(\bar\chi\chi)(\bar f f)}{M_*^2} \,,
    & \text{($s$-channel, scalar)} \label{eq:2} \\
  \mathcal{O}_A &= \frac{(\bar\chi\gamma_\mu\gamma_5\chi)(\bar f \gamma^\mu\gamma_5 f)}{M_*^2} \,,
    & \text{($s$-channel, axial vector)} \label{eq:3} \\
  \mathcal{O}_t &= \frac{(\bar\chi f)(\bar f \chi)}{M_*^2} \,.
    & \text{($t$-channel, scalar)} \label{eq:4} 
\end{align}

While this set of operators is not exhaustive, it encompasses the essential phenomenologically distinct scenarios: spin-dependent and spin-independent dark matter--nucleus scattering, as well as $s$- and $p$-wave annihilation. The classification of the effective operators as $s$-channel or $t$-channel refers to the renormalizable model from which they typically arise: \eqref{eq:1}--\eqref{eq:3} are most straightforwardly obtained if dark matter pair production is mediated by a new neutral particle propagating in the $s$-channel, while Eq.~\eqref{eq:4} arises naturally if the mediator is a charged scalar exchanged in the $t$-channel (for instance a squark or slepton). The suppression scale $M_*$ can be interpreted as the mass of the mediator $M$, divided by the geometric mean of its couplings to SM fermions, $g_f$, and dark matter, $g_\chi$: $M_* = M / \sqrt{g_f g_\chi}$. It is worth noting that the DM particles are not explicitly assumed to interact via the weak force; they may as well couple to the SM through a new force. Within this framework, interactions of SM particles and WIMPs are described by only two parameters, the suppression scale $M_*$ and the WIMP mass \mx. 

Previous work relating collider searches to direct and indirect searches for dark matter has focused on the Tevatron~\cite{cdf11,cdf12,cdf2} and the LEP~\cite{delphi1,delphi2}. While the hadronic machine probes the dark matter couplings to light quarks, the LEP data are sensitive to the DM--electron coupling. In general, Tevatron constraints are very strong for lighter dark matter and fall off when the dark matter mass exceeds the typical energy reach of the collider~\cite{dm-lhc}. The constraints also depend on the coupling of the dark matter; if the dark matter primarily couples to gluons, the constraints from colliders become especially strong~\cite{tevatron1}. Moreover, if the possibility of light mediators is taken into account, as motivated by cosmic-ray excesses~\cite{cosmic-excess1,cosmic-excess2}, the introduction of a light mediator of mass $\lesssim 10~\gev$ alleviates the monojet bounds completely for most cases. This leads to the conclusion that if a direct dark matter signal is established in a region that is in conflict with collider bounds, a new light state should be introduced to reconcile the data~\cite{tevatron2}. One mode in which dark matter may be searched for at LEP, with relatively little model dependence, is its pair production in association with a hard photon. The LEP experiments have searched for anomalous monophoton events in their data sets, but have found no discrepancy from the prediction of the standard model~\cite{lep}.

%%%%%%%%%%%%%%%%%%%%%%%%%%%%%%%%%%%%%%%%%%%%%%%%%%%%%%%%%%%%%%%%%%%%%%%%%%%%%%%%%%%%%%%%%%%%%%%%%%%%%%%%%%%%
\subsection{Monojet searches}\label{sc:monojet}
%%%%%%%%%%%%%%%%%%%%%%%%%%%%%%%%%%%%%%%%%%%%%%%%%%%%%%%%%%%%%%%%%%%%%%%%%%%%%%%%%%%%%%%%%%%%%%%%%%%%%%%%%%%%

Event topologies with a single high-\et\ jet and large \met, henceforth referred to as \emph{monojets}, are important probes of physics beyond the SM at the LHC. The ATLAS~\cite{atlas-monojet} and CMS~\cite{cms-monojet} experiments have performed searches for an excess of monojet events over SM expectations. The analyses outlined here use the full 2011 $pp$ LHC dataset at a center-of-mass energy of $\sqrt{s} = 7~\tev$. The primary SM process yielding a true monojet final state is $Z$-boson production in association with a jet, where $Z\to\nu\nu$. Other known processes acting as background in this search are $Z(\to\ell\ell)$+jets, \W+jets, $t\bar{t}$ and single-top events, with $\ell=e,\,\mu$. All electroweak backgrounds and multijet events passing the selections criteria, as well as non-collision backrounds, are determined by data-driven methods. Top and diboson backgrounds are determined solely from Monte Carlo (MC) simulation.  

The monojet analyses for ATLAS and CMS are based on some more-or-less common requirements: large \met, with thresholds typically ranging from 120~\gev\ to 500~\gev\ and a energetic jet with a variable \pt\ threshold higher than 110~\gev\ that fulfills high jet-reconstruction quality criteria. In addition, events with at least one electron or muon or a third jet are rejected. Back-to-back dijet events are suppressed by requiring the subleading jet not to point in the direction of $\bf p_{\bf T}^{\bf miss}$. The selected data are required to pass a trigger based on high \met\ (ATLAS) or large \met\ plus one high-\et\ jet (CMS).  
   
The data, amounting to $\sim5~\ifb$, are found to be in agreement with the SM expectations. The results are interpreted in a framework of WIMP production with the simulated WIMP-signal MC samples corresponding to various assumptions of the effective field theory, as discussed previously; some of the options are listed in Table~\ref{tb:operators}. The derived limits are independent of the theory behind the WIMP (SUSY, extra dimensions, etc), yet it \emph{has} been assumed that other hypothetical particles are two heavy to be produced directly in $pp$ collisions. In the presented limits, a Dirac DM fermion is considered, however conclusions for Majorana fermions can also be drawn, since their production cross section only differs by a factor of two. In this framework, the interaction between SM and DM particles are defined by only two parameters, namely the DM-particle mass, \mx, and the suppression scale, $M_*$, which is related to the mediator mass and to its coupling to SM and DM particles.

\begin{table}[htb]
\tbl{Effective interaction operators of WIMP pair production considered in the monojet and monophoton analyses, following the formalism of Ref.~\citen{tevatron1}.}
{\begin{tabular}{@{}cccc@{}} \toprule
Name & Initial state & Type & Operator \\ \colrule
D1  & $qq$ & scalar       & $\frac{m_q}{M_*^3}\bar{\x}\x\bar{q}q$ \\
D5  & $qq$ & vector       & $\frac{1}{M_*^2}\bar{\x}\gamma^{\mu}\x\bar{q}\gamma_{\mu}q$ \\
D8  & $qq$ & axial-vector & $\frac{1}{M_*^2}\bar{\x}\gamma^{\mu}\gamma^5\x\bar{q}\gamma_{\mu}\gamma^{\mu}q$ \\
D9  & $qq$ & tensor       & $\frac{1}{M_*^2}\bar{\x}\sigma^{\mu\nu}\x\bar{q}\sigma_{\mu\nu}q$ \\
D11 & $gg$ & scalar       & $\frac{1}{4M_*^3}\bar{\x}\x\alpha_s(G^s_{\mu\nu})^2$\\ \botrule
\end{tabular} \label{tb:operators}}
\end{table}

Experimental and theoretical systematic uncertainties are considered when setting limits on the model parameters $M_*$ and \mx. The experimental uncertainties on jet energy scale and resolution and on \met\ range from $1-20\%$ of the WIMP event yield, depending on the \met\ and \pt\ thresholds and the considered interaction operator. Other experimental uncertainties include the ones associated with the trigger efficiency and the luminosity measurement. On the other hand, the parton-distribution-function set, the amount of ISR/FSR, and the factorization and renormalization scales assumed lead to theoretical uncertainties on the simulated WIMP signal. 

From the limit on the visible cross section of new physics processes BSM, lower limits on the suppression scale as a function of the WIMP mass have been derived by the ATLAS Collaboration~\cite{atlas-monojet}. The 90\% confidence level (CL) lower limits for the D5 and D11 operators are shown in Fig.~\ref{fg:atlas-monojet-d}. The observed limit on $M_*$ includes experimental uncertainties; the effect of theoretical uncertainties is indicated by dotted $\pm1\sigma$ lines above and below it. Around the expected limit, $\pm1\sigma$ variations due to statistical and systematic uncertainties are shown as a gray band. The lower limits are flat up to $\mx\lesssim100~\gev$ and become weaker at higher mass due to the collision energy. In the bottom-right corner of the $\mx-M_*$ plane (light-gray shaded area), the effective field theory approach is no longer valid. The rising lines correspond to couplings consistent with the measured thermal relic density~\cite{tevatron1}, assuming annihilation in the early universe proceeded exclusively via the given operator. Similar exclusion limits for all operators listed in Table~\ref{tb:operators} are given in Ref.~\citen{atlas-monojet}. For the operator D1, the limits are much weaker ($\sim30~\gev$) than for other operators. Nevertheless, if heavy-quark loops are included in the analysis, much stronger bounds on $M_*$ can be obtained~\cite{monojet-loop}.

\begin{figure}[htb]
\centerline{\includegraphics[width=0.49\textwidth]{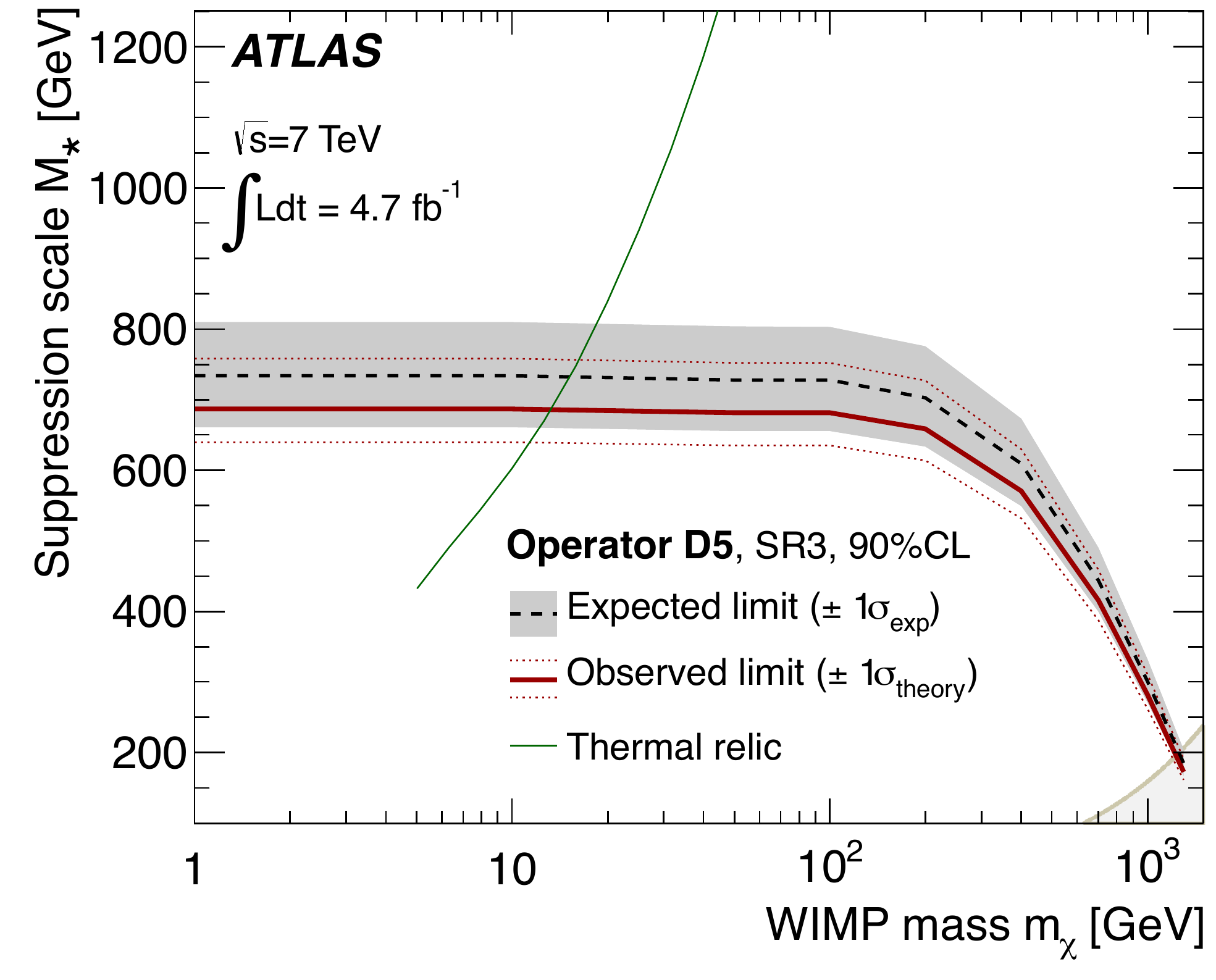}
  \hspace{0.02\textwidth}
  \includegraphics[width=0.49\textwidth]{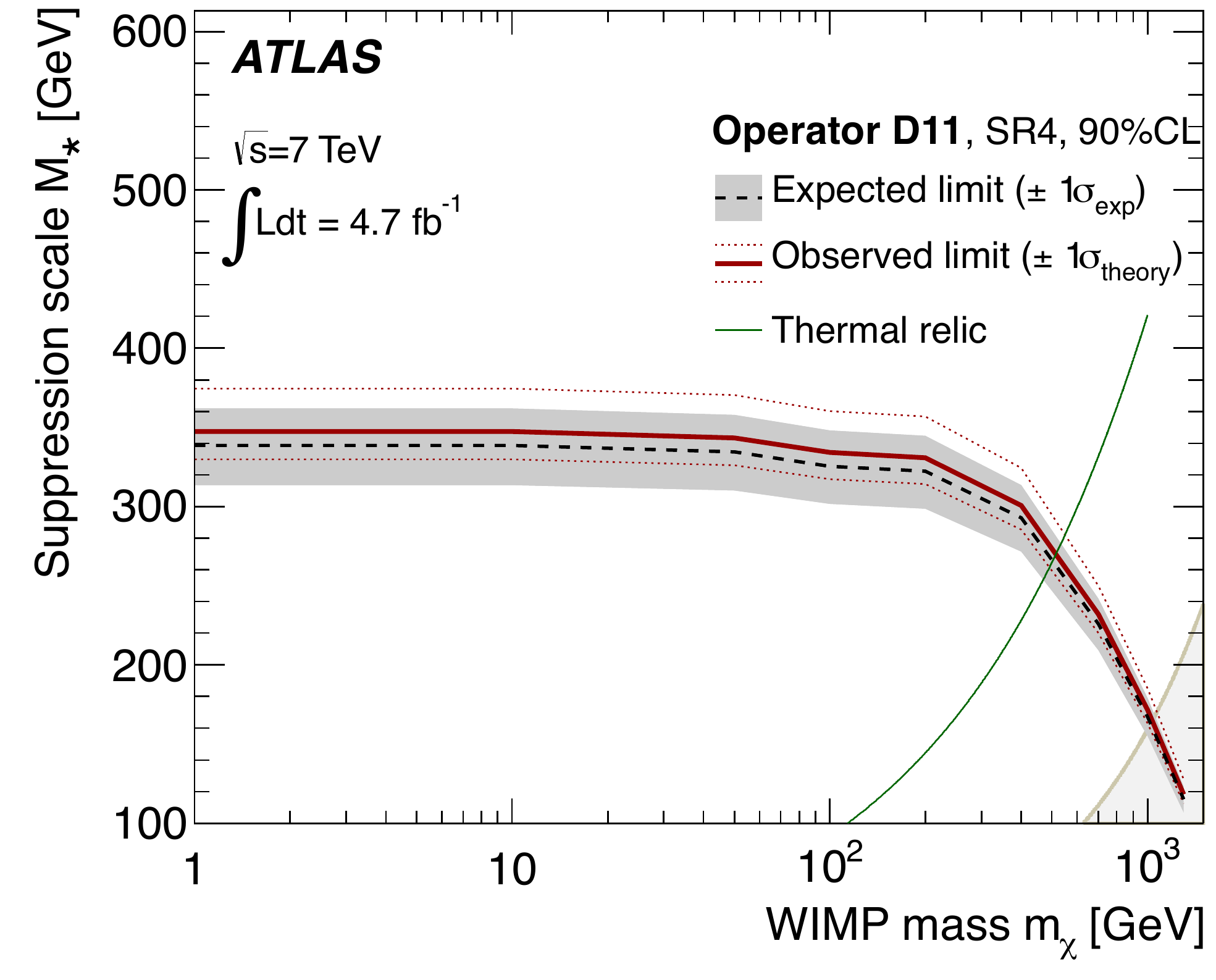}} 
\caption{ATLAS lower limits at 90\% CL on the suppression scale, $M_*$, for different masses of \x\ obtained with the monojet analysis. The region below the limit lines is excluded. All shown curves and areas are explained in the text. From Ref.~\citen{atlas-monojet}.\label{fg:atlas-monojet-d}}
\end{figure}

The observed limit on the dark matter--nucleon scattering cross section depends on the mass of the dark matter particle and the nature of its interaction with the SM particles. The limits on the suppression scale as a function of \mx\ can be translated into a limit on the cross section using the reduced mass of \x--nucleon system~\cite{tevatron2}, which can be compared with the constraints from direct and indirect detection experiments. Figure~\ref{fg:monojet-si-sd} shows the 90\% CL upper limits on the dark matter--nucleon scattering cross section as a function of the mass of DM particle for the spin-independent (left) and spin-dependent (right) models obtained by CMS~\cite{cms-monojet}. Limits from CDF~\cite{cdf2}, XENON100~\cite{xenon100a}, CoGent~\cite{cogent}, CDMS~II~\cite{cdmsii1,cdmsii2}, SIMPLE~\cite{simple}, COUPP~\cite{coupp1}, Picasso~\cite{picasso1}, IceCube~\cite{icecube1}, Super-K~\cite{superk}, as well as the CMS monophoton~\cite{cms-monophoton} analysis are superimposed for comparison. Similar limits have been obtained by the ATLAS experiment~\cite{atlas-monojet}.

\begin{figure}[htb]
\centerline{\includegraphics[width=0.5\textwidth]{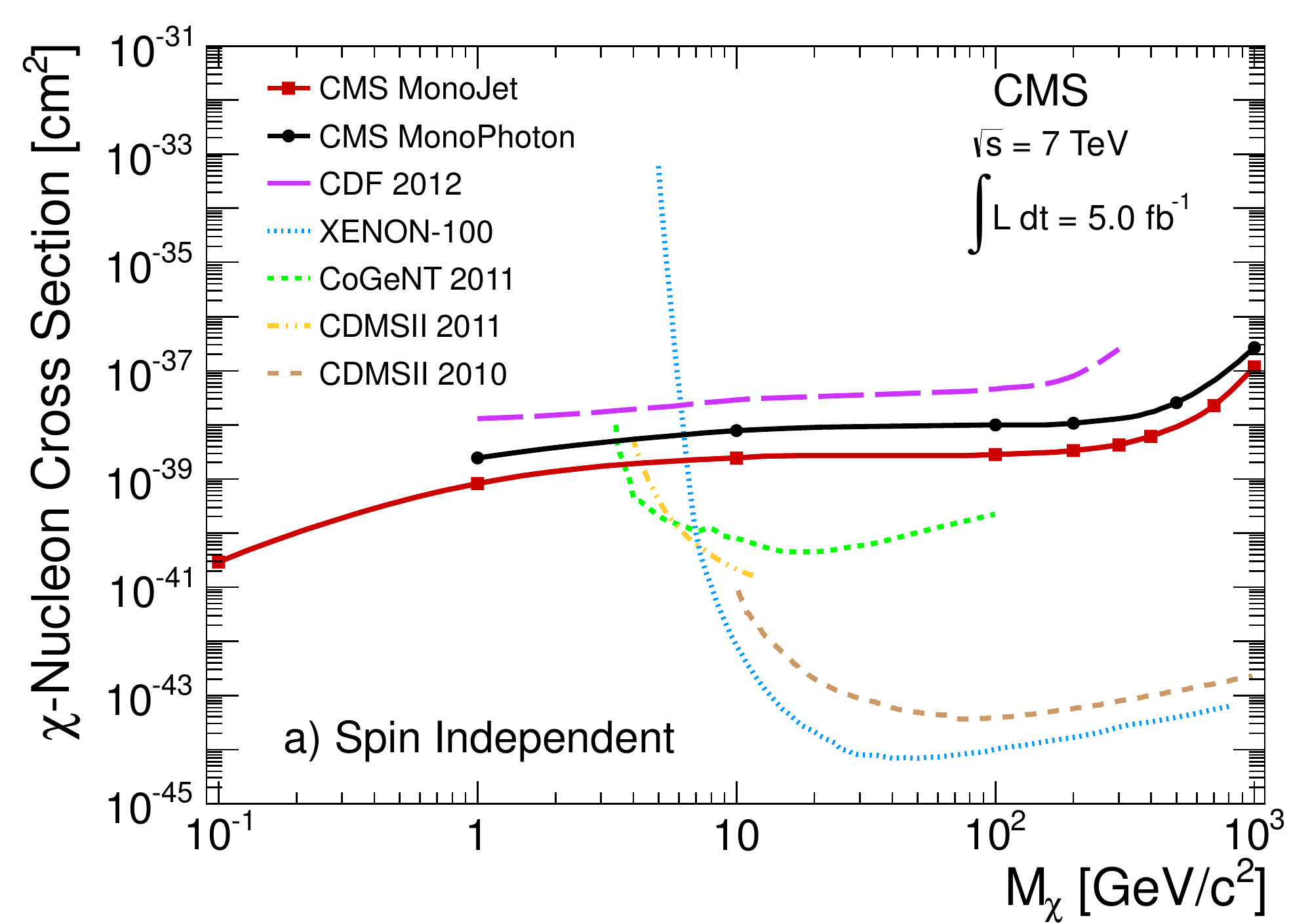}
  \hspace{0.0\textwidth}
  \includegraphics[width=0.5\textwidth]{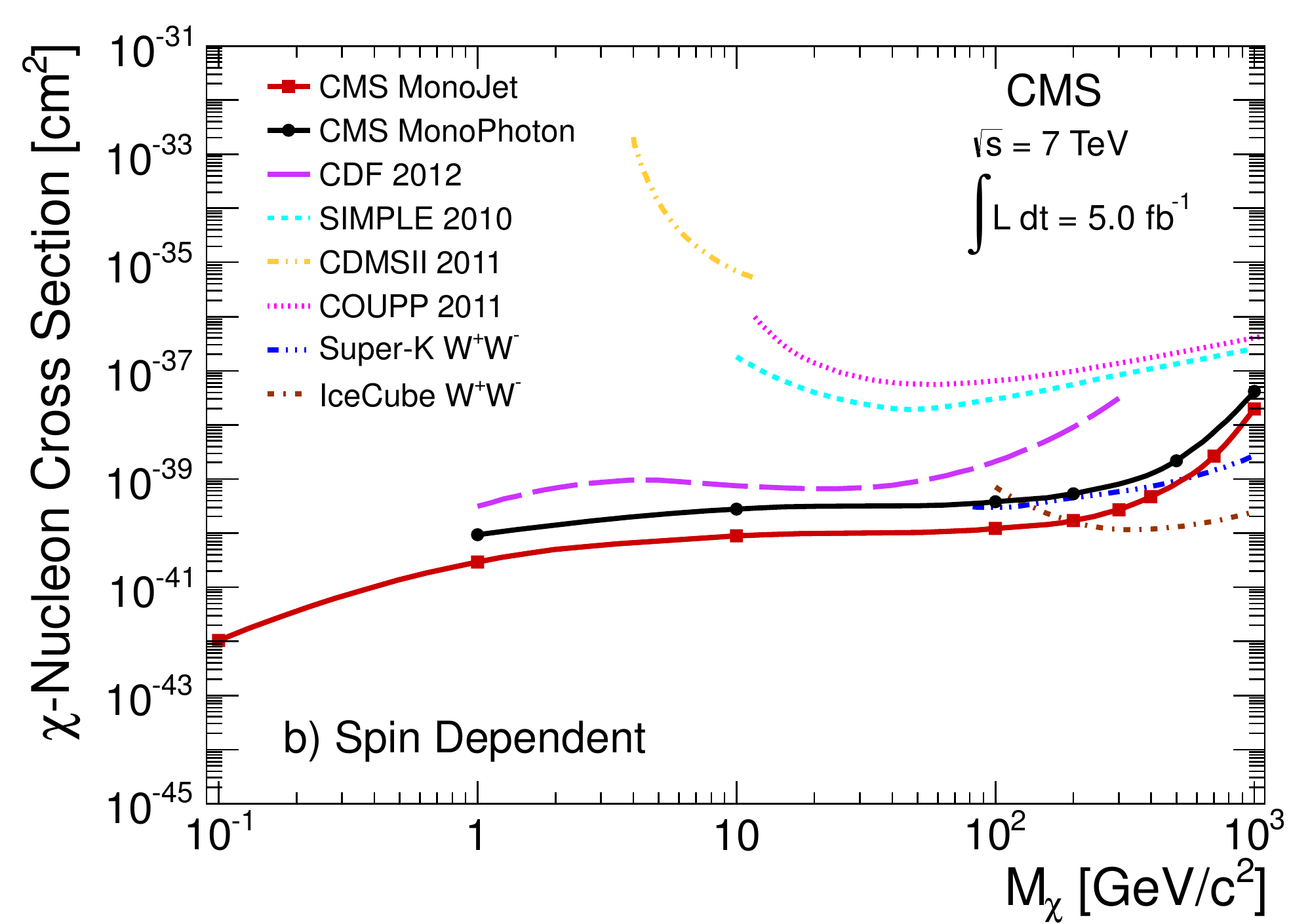}} 
\caption{90\% CL upper limits on the dark matter--nucleon scattering cross section versus DM particle mass for the spin-independent (left) and spin-dependent (right) obtained with the CMS monojet analysis. Explanation of the shown curves is given in the text. From Ref.~\citen{cms-monojet}.\label{fg:monojet-si-sd}}
\end{figure}

The spin-dependent limits, derived from the operator D8, give a smaller, hence better, bound on the WIMP-nucleon cross section throughout the range of \mx, compared to direct DM experiments. In the spin-independent case the bounds from direct detection experiments are stronger for $\mx\gtrsim10~\gev$, whereas the collider bounds, acquired with the operator D5, get important for the region of low DM masses. 

The ATLAS collider limits on vector (D5) and axial-vector (D8) interactions are also interpreted in terms of the relic abundance of WIMPs, using the same effective theory approach~\cite{tevatron1}. The upper limits on the annihilation rate of WIMPs into light quarks, defined as the product of the annihilation cross section $\sigma$ and the relative WIMP velocity $v$ averaged over the WIMP velocity distribution $\langle\sigma v\rangle$, are shown in Fig.~\ref{fg:atlas-monojet-annih}. The results are compared to limits on WIMP annihilation to $b\bar{b}$, obtained from galactic high-energy gamma-ray observations, measured by the Fermi-LAT telescope~\cite{fermilat}. Gamma-ray spectra and yields from WIMPs annihilating to $b\bar{b}$, where photons are produced in the hadronization of the quarks, are expected to be very similar to those from WIMPs annihilating to light quarks~\cite{annihilation1,annihilation2}. Under this assumption, the ATLAS and Fermi-LAT limits can be compared, after scaling up the Fermi-LAT values by a factor of two to account for the Majorana-to-Dirac fermion adaptation. Again, the ATLAS bounds are especially important for small WIMP masses: below 10~\gev\ for vector couplings and below about $100~\gev$ for axial-vector ones. In this region, the ATLAS limits are below the annihilation cross section needed to be consistent with the thermic relic value, keeping the assumption that WIMPs have annihilated to SM quarks only via the particular operator in question. For masses of $\mx\gtrsim 200~\gev$ the ATLAS sensitivity becomes worse than the Fermi-LAT one. In this region, improvements can be expected when going to larger center-of-mass energies at the LHC.

\begin{figure}[htb]
\centerline{\includegraphics[width=0.6\textwidth]{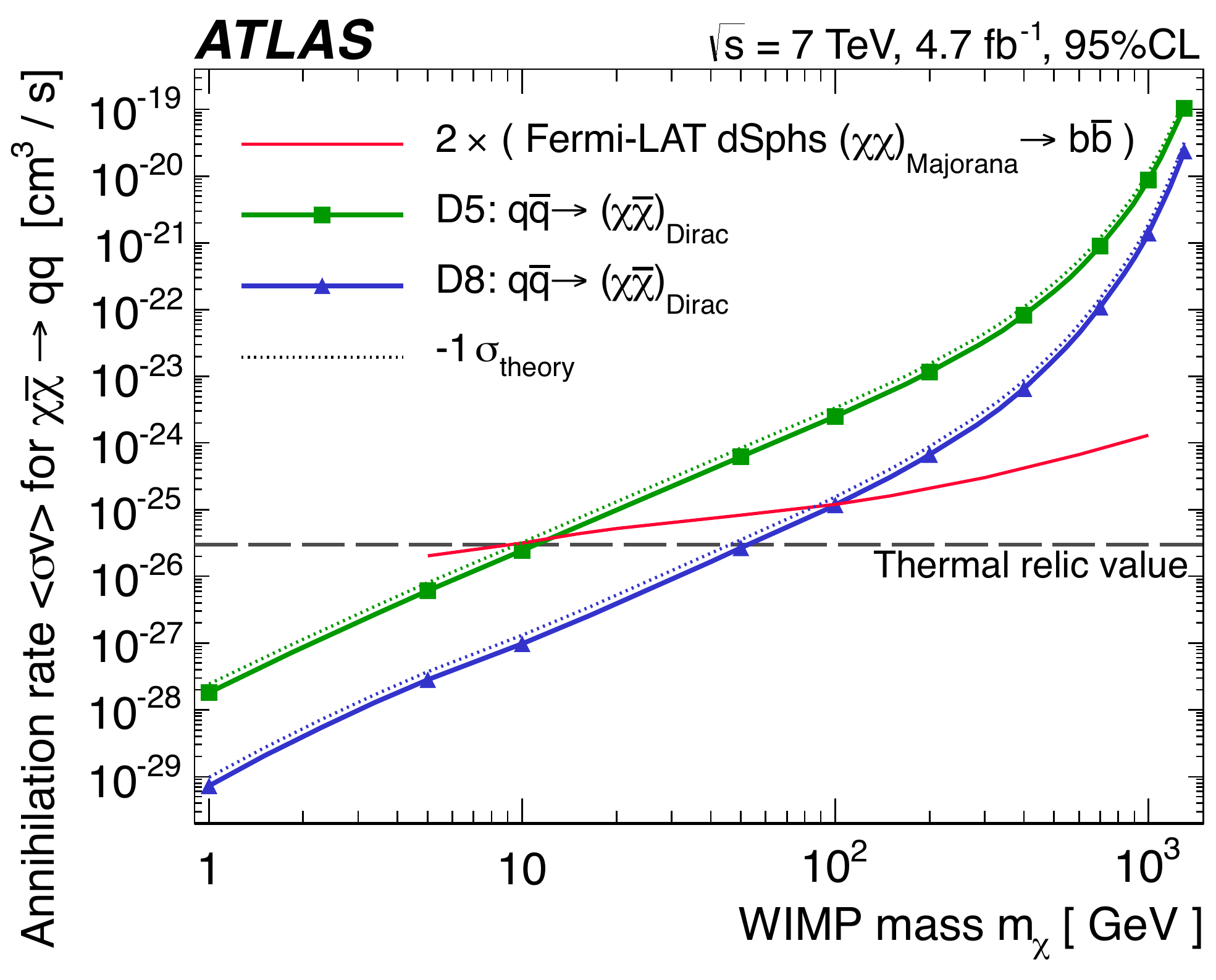}}
\caption{Inferred ATLAS 95\% CL limits on WIMP annihilation rates $\langle\sigma v\rangle$ versus mass \mx. 
Explanation of the shown curves is given in the text. From Ref.~\citen{atlas-monojet}.\label{fg:atlas-monojet-annih}}
\end{figure}

%%%%%%%%%%%%%%%%%%%%%%%%%%%%%%%%%%%%%%%%%%%%%%%%%%%%%%%%%%%%%%%%%%%%%%%%%%%%%%%%%%%%%%%%%%%%%%%%%%%%%%%%%%%%
\subsection{Monophoton-based probes}\label{sc:monophoton}
%%%%%%%%%%%%%%%%%%%%%%%%%%%%%%%%%%%%%%%%%%%%%%%%%%%%%%%%%%%%%%%%%%%%%%%%%%%%%%%%%%%%%%%%%%%%%%%%%%%%%%%%%%%%

Similarly to the monojet searches, the \emph{monophoton} analyses aim at probing dark matter requiring large \met\ ---from the \x-pair production--- and at least one ISR/FSR photon. Searches in monophoton events by ATLAS~\cite{atlas-monophoton} and CMS~\cite{cms-monophoton} also show an agreement with the SM expectations. The limits are derived in a similar fashion as for monojet search, however the monophoton search is found to be somewhat less sensitive with respect to the monojet topology.

The primary background for a $\gamma+\met$ signal is the irreducible SM background from $Z\gamma\to\nu\bar{\nu}\gamma$ production. This and other SM backgrounds, including $W\gamma$, $W\to e\nu$, $\gamma+\text{jet}$ multijet, diphoton and diboson events, as well as backgrounds from beam halo and cosmic-ray muons, are taken into account in the analyses. The CMS analysis is based on singe-photon triggers, whilst ATLAS relies on high-\met\ triggered events. The photon candidate is required to pass tight quality and isolation criteria, in particular in order to reject events with electrons faking photons. The missing transverse momentum of the selected events should be as high as 150~\gev\ (130~\gev) in the ATLAS (CMS) search. In CMS, events with a reconstructed jet are vetoed, while the ATLAS analysis rejects events with an electron, a muon or a second jet. 

Both analyses, observe no significant excess of events over the expected background when applied on $\sim5~\ifb$ of $pp$ collision data at $\sqrt{s}=7~\tev$. Hence they set lower limits on the suppression scale, $M_*$ versus the DM fermion mass, \mx, which in turn they are translated into upper limits on the nucleon--WIMP interaction cross section applying the prescription in Ref.~\citen{tevatron1}. Figure~\ref{fg:atlas-monophoton} shows the 90\% CL upper limits on the nucleon--WIMP cross section as a function of \mx\ derived from the ATLAS search~\cite{atlas-monophoton}. The results are compared with previous CDF~\cite{cdf2}, CMS~\cite{cms-monojet,cms-monophoton} and direct WIMP detection experiments~\cite{xenon100a,cdmsii1,cdmsii2,cogent,simple,picasso2} results. The CMS limit curve generally overlaps the ATLAS curve.

\begin{figure}[htb]
\centerline{\includegraphics[width=0.85\textwidth]{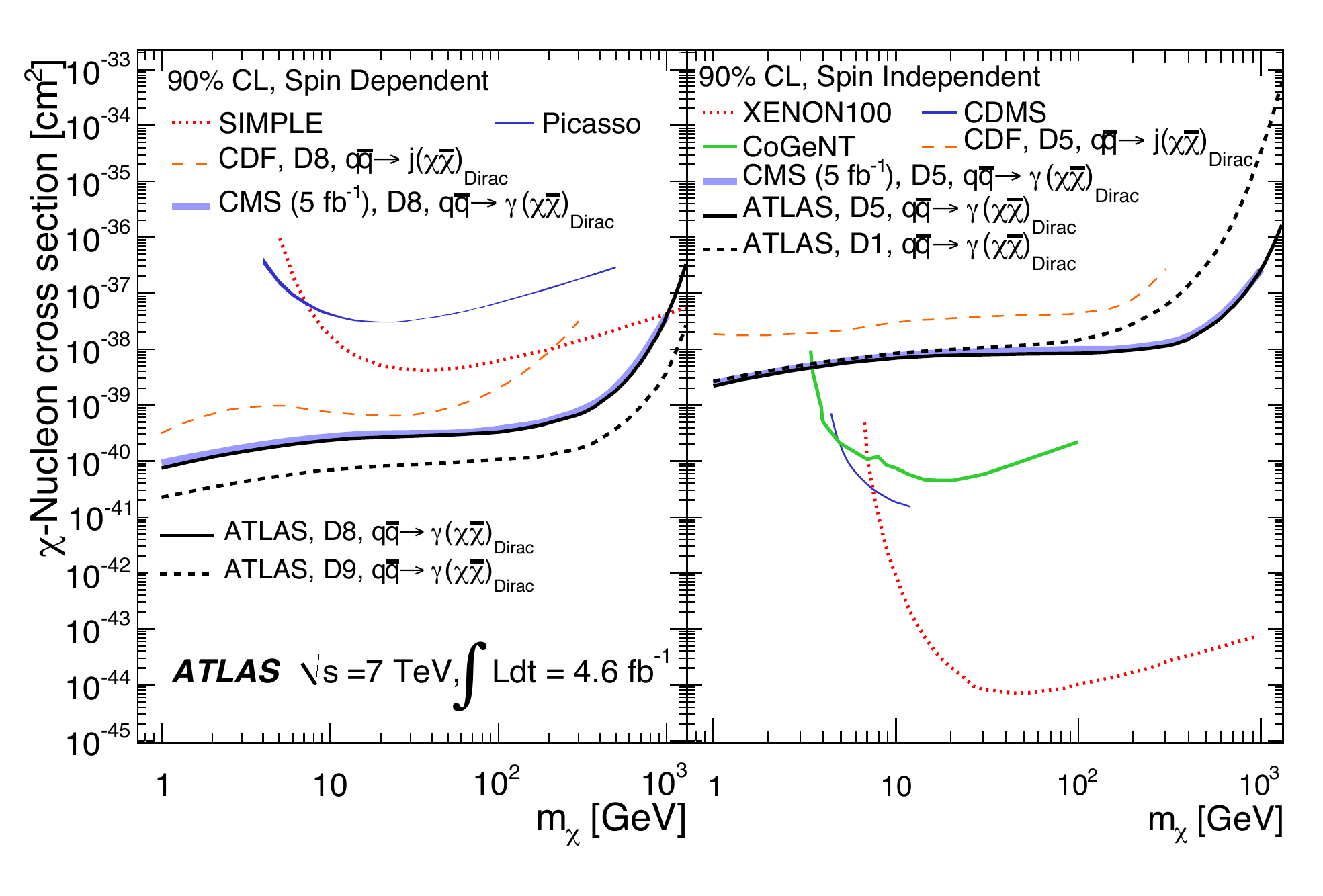}}
\caption{ATLAS-derived 90\% CL upper limits on the nucleon-WIMP cross section as a function of \mx\ for spin-dependent (left) and spin-independent (right) interactions, corresponding to D8, D9, D1, and D5 operators.   
Explanation of the shown curves is given in the text. From Ref.~\citen{atlas-monophoton}.\label{fg:atlas-monophoton}}
\end{figure}

The observed limits on $M_*$ typically decrease by 2\% to 10\% if the $-1\sigma$ theoretical uncertainty, resulting from the same sources as the one cited in the monojet analysis, is considered. This translates into a 10\% to 50\% increase of the quoted nucleon-WIMP cross section limits. To recapitulate, the exclusion in the region $1~\gev<\mx<1~\gev$ ($1~\gev<\mx<3.5~\gev$) for spin-dependent (spin-independent) nucleon--WIMP interactions is driven by the results from collider experiments, always under the assumption of the validity of the effective theory, and is still dominated by the monojet results.

%%%%%%%%%%%%%%%%%%%%%%%%%%%%%%%%%%%%%%%%%%%%%%%%%%%%%%%%%%%%%%%%%%%%%%%%%%%%%%%%%%%%%%%%%%%%%%%%%%%%%%%%%%%%
\subsection{Mono-$W$ and mono-$Z$ final states}\label{sc:monoWZ}
%%%%%%%%%%%%%%%%%%%%%%%%%%%%%%%%%%%%%%%%%%%%%%%%%%%%%%%%%%%%%%%%%%%%%%%%%%%%%%%%%%%%%%%%%%%%%%%%%%%%%%%%%%%%

As demonstrated in the previous sections, searches for monojet or monophoton signatures have yielded powerful constraints on dark matter interactions with SM particles. Other studies propose probing DM at LHC through a  
$pp\to\x\bar{\x}+W/Z$, with a leptonically decaying $W$~\cite{monow} or $Z$~\cite{monoz1,monoz2}. The final state in this case would be large \met\ and a single charged lepton (electron or muon) for the \emph{mono-W} signature (\emph{monolepton}) or a pair of charged leptons that reconstruct to the $Z$ mass for the \emph{mono-Z} signature. In either case, the gauge boson radiations off a $q\bar{q}$ initial state and an effective field theory is deployed to describe the  contact interactions that couple the SM particle with the WIMP.  

In Ref.~\citen{monow}, the existing $W'$ searches from CMS~\cite{cms-wprime7,cms-wprime8} ---which share a similar final state with mono-$W$ searches--- are used to place a bound on mono-$W$ production at LHC, which for some choices of couplings are better than monojet bounds. This is illustrated in the left panel of Fig.~\ref{fg:monow-monoz}, where the spin-independent WIMP-proton cross section limits are drawn. The parameter $\xi$ parametrizes the relative strength of the coupling to down-quarks with respect to up-quarks: $\xi=+1$ for equal couplings; $\xi=-1$ for opposite-sign ones; and $\xi=0$ when there is no coupling to down quarks. Even in cases where the monoleptons do not provide the most stringent constraints, they provide an interesting mechanism to disentangle WIMP couplings to up-type versus down-type quarks. This analysis has been followed up by CMS~\cite{cms-monow} with the full 2012 dataset yielding similar to limits based on $W'$ searches.  

\begin{figure}[htb]
\centerline{\includegraphics[width=0.5\textwidth]{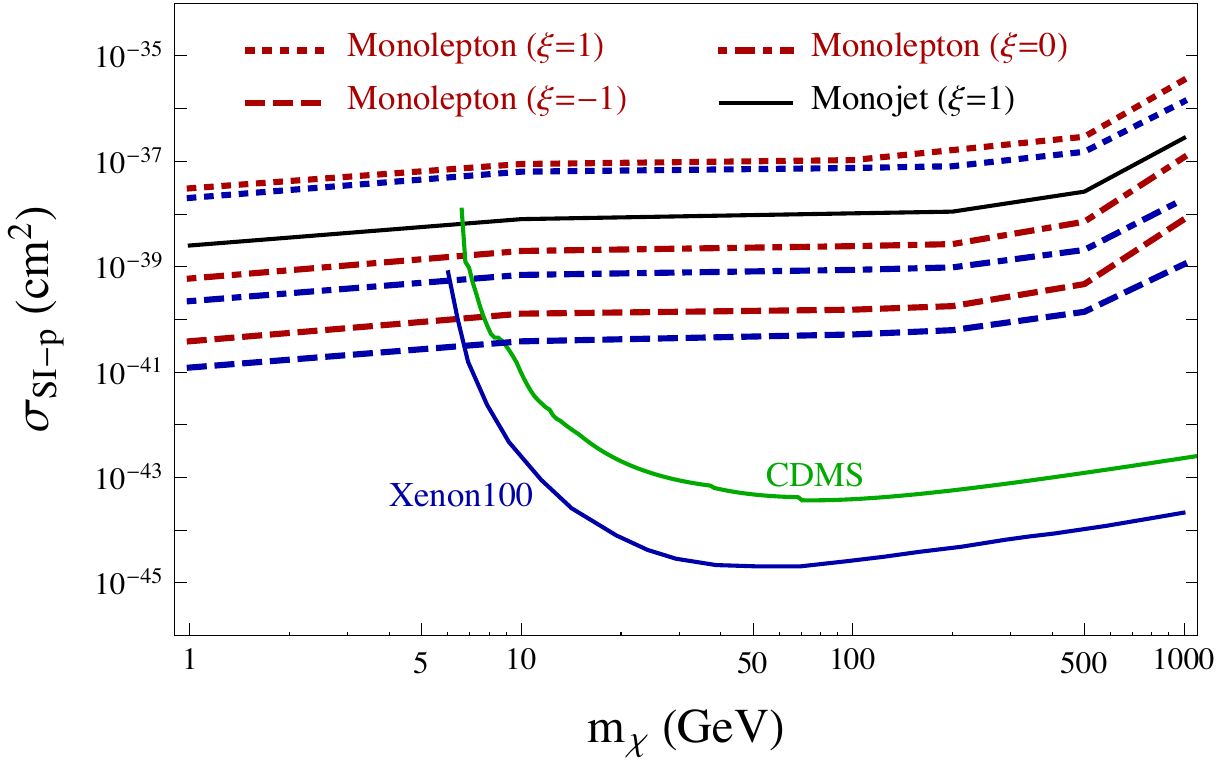}
  \hspace{0.0\textwidth}
  \includegraphics[width=0.5\textwidth]{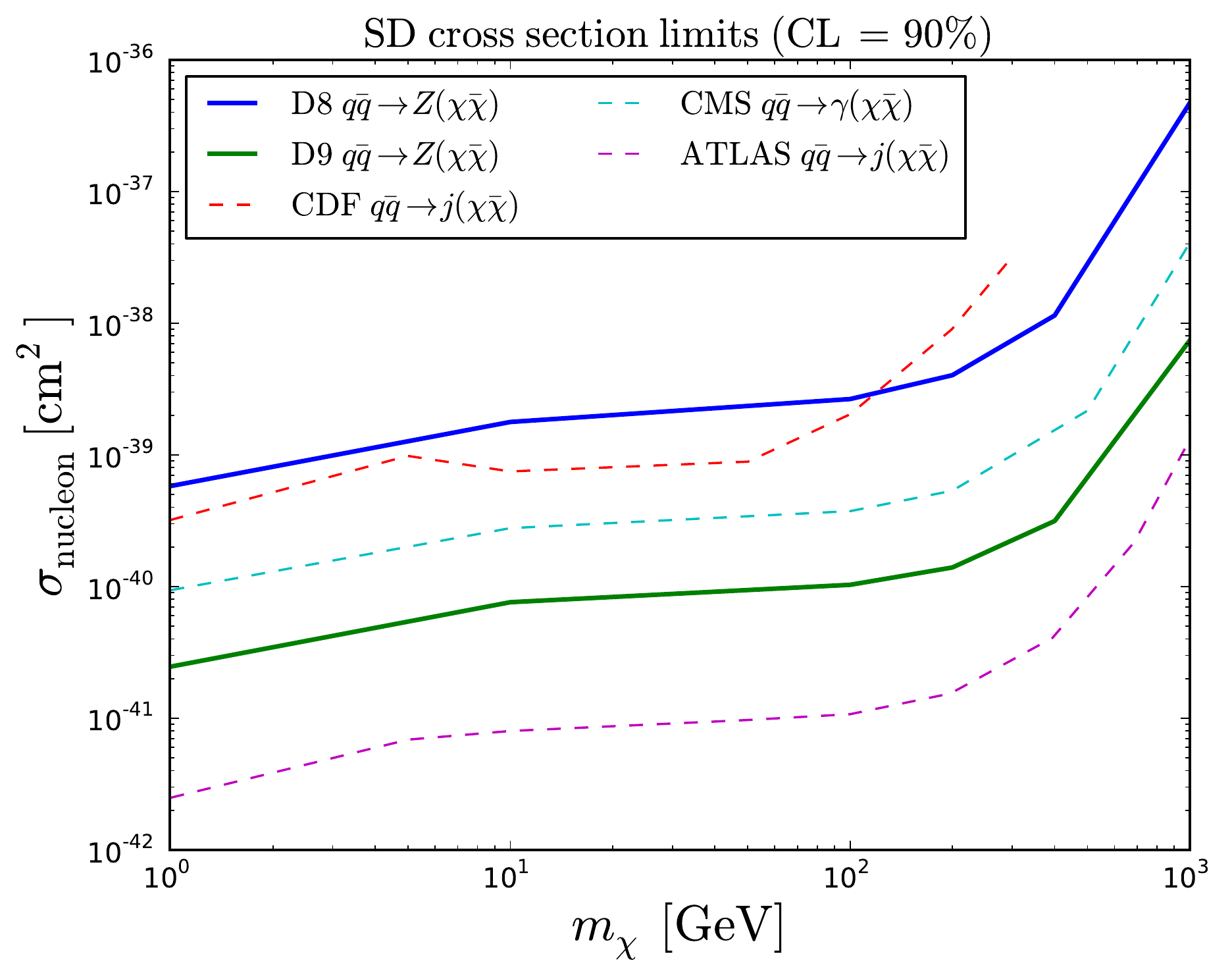}} 
\caption{\emph{Left:} Monolepton bounds and bounds from direct detection projected into the plane of the WIMP mass and the spin-independent cross section with protons. The red and blue lines are from CMS $W'$ searches: $5~\ifb$ data~\cite{cms-wprime7} at 7~\tev\ and $20~\ifb$ data~\cite{cms-wprime8} at 8~\tev, respectively. From Ref.~\citen{monow}.
\emph{Right:} Mono-$Z$ exclusion regions at 90\% CL in the DM mass versus WIMP--nucleon spin-dependent cross section plane obtained from the ATLAS $ZZ$ measurement~\cite{atlas-zz}. Existing collider limits are also shown for comparison. From Ref.~\citen{monoz2}.\label{fg:monow-monoz}}
\end{figure}

Furthermore a (leptonic) mono-$Z$ signal has been considered in the literature, highlighting the kinematic features~\cite{monoz1} and recasting LHC results to constrain models~\cite{monoz2}. Specifically, the ATLAS measurement~\cite{atlas-zz} of $ZZ\to\ell\ell\nu\nu$ carried out with $\sim5~\ifb$ of 7~\tev\ data has been reinterpreted into a bound on production of dark matter in association with a $Z$ boson. The obtained bounds for the spin-dependent WIMP--nucleon cross section is depicted in the right panel of Fig.~\ref{fg:monow-monoz} along with other bounds from colliders~\cite{cdf2,cms-monojet,atlas-monojet}. The mono-$Z$ signature yields limits which are somewhat weaker than those from monojets or monophotons. Nevertheless, leptonic mono-$Z$ searches are less subject to systematic uncertainties from jet energy scales and photon identification, and hence may scale better at large integrated luminosities.
 
The ATLAS Collaboration has recently~\cite{atlas-monowz} extended the range of possible mono-$X$ probes by looking for $pp\to\x\bar{\x}+W/Z$, when the gauge boson decays to two quarks, as opposed to the leptonic signatures discussed so far. The analysis searches for the production of $W$ or $Z$ bosons decaying hadronically and reconstructed as a single massive jet in association with large \met\ from the undetected $\x\bar{\x}$ particles. For this analysis, the jet candidates are reconstructed using a filtering procedure referred to as \emph{large-radius jets}~\cite{fat-jets}. This search, the first of its kind, is sensitive to WIMP pair production, as well as to other DM-related models, such as invisible Higgs boson decays ($WH$ or $ZH$ production with $H\to\x\bar{\x}$).

\begin{figure}[htb]
\centerline{\includegraphics[width=0.8\textwidth]{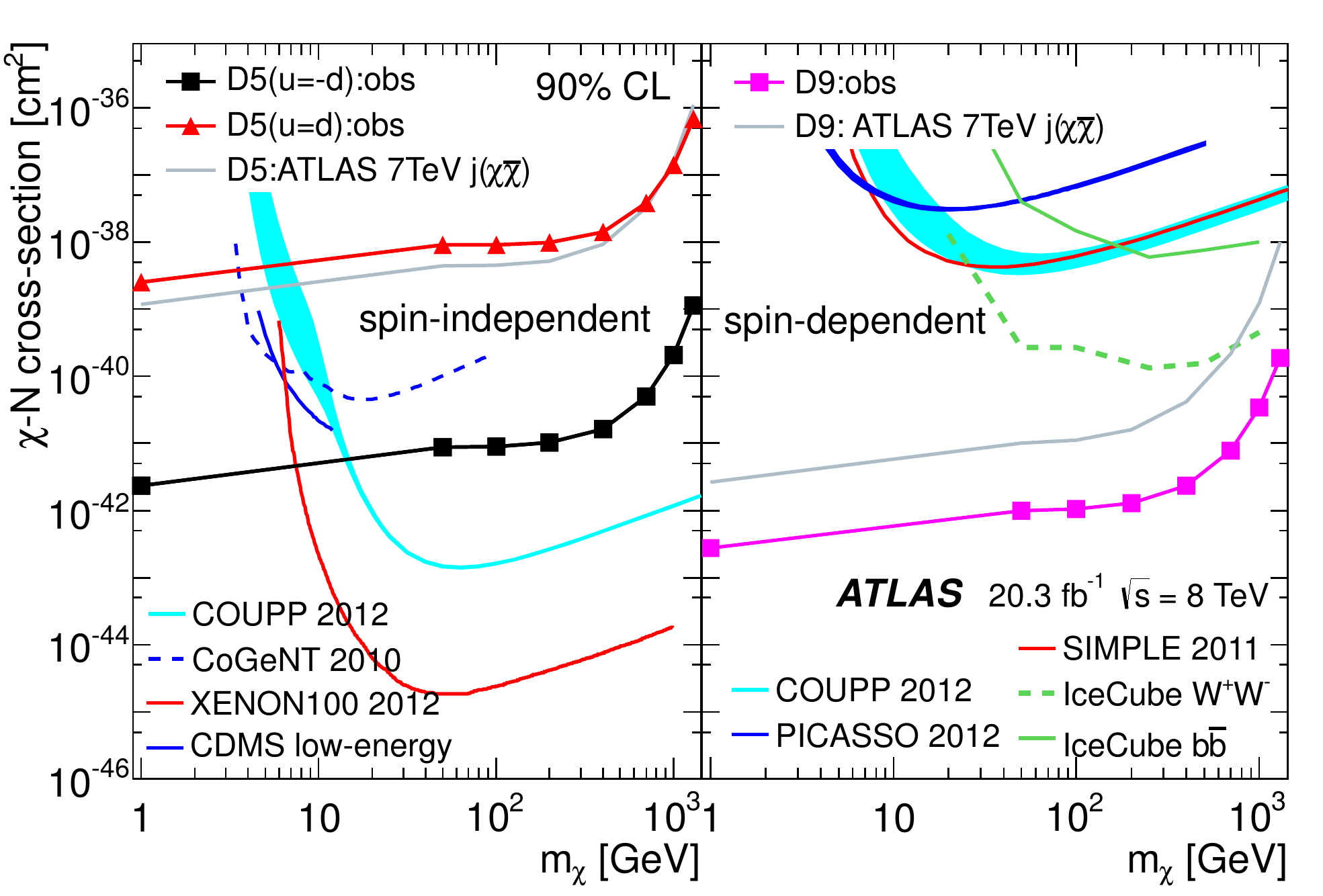}}
\caption{ATLAS-derived limits on \x--nucleon cross sections as a function of \mx\ at 90\% CL for spin-independent (left) and spin-dependent (right) cases, obtained with the mono-$W/Z$ analysis and compared to previous limits. From Ref.~\citen{atlas-monowz}.\label{fg:atlas-monowz}}
\end{figure}

As shown in Fig.~\ref{fg:atlas-monowz}, this search for dark matter pair production in association with a $W$ or $Z$ boson extends the limits on the dark matter--nucleon scattering cross section in the low mass region $\mx<10~\gev$ where the direct detection experiments have less sensitivity. The new limits are also compared to the limits set by ATLAS in the $7~\tev$ monojet analysis~\cite{atlas-monojet} and by direct detection experiments~\cite{xenon100b,cdmsii1,cdmsii2,cogent,picasso2,simple,icecube2,coupp2}. For the spin-independent case with the opposite-sign up-type and down-type couplings, the limits are improved by about three orders of magnitude. For other cases, the limits are similar.

To complement the effective-field-theory models, limits are calculated for an UV-complete theory with a light mediator, the Higgs boson. The upper limit on the cross section of the Higgs boson production through $WH$ and $ZH$ modes and decay to invisible particles is 1.3~pb at 95\% CL for $m_H=125~\gev$. Figure~\ref{fg:atlas-monowz-h} shows the upper limit of the total cross section of $WH$ and $ZH$ processes with $H\to\x\bar{\x}$, normalized to the SM next-to-leading order prediction for the $WH$ and $ZH$ production cross section (0.8~pb for $m_H=125~\gev$)~\cite{higgsnlo}, which is 1.6 at $95\%$ CL for $m_H=125$ GeV.

\begin{figure}[htb]
\centerline{\includegraphics[width=0.65\textwidth]{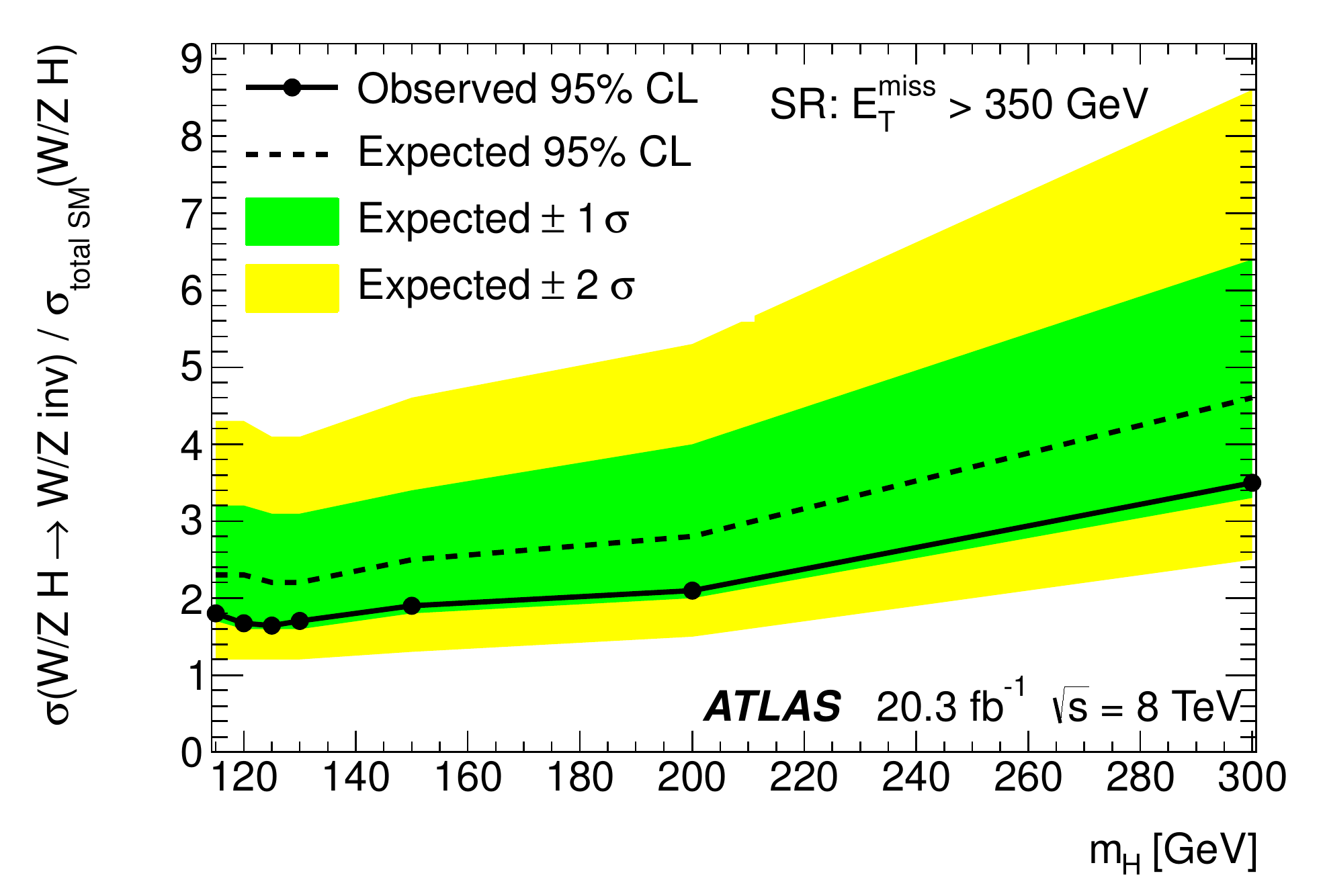}}
\caption{ATLAS-derived limit on the Higgs boson cross section for decay to invisible particles divided by the cross section for decays to Standard Model particles as a function of $m_H$ at 95\% CL, derived from the mono-$W/Z$ analysis with $\met > 350~\gev$. From Ref.~\citen{atlas-monowz}.\label{fg:atlas-monowz-h}}
\end{figure}

%%%%%%%%%%%%%%%%%%%%%%%%%%%%%%%%%%%%%%%%%%%%%%%%%%%%%%%%%%%%%%%%%%%%%%%%%%%%%%%%%%%%%%%%%%%%%%%%%%%%%%%%%%%%
\subsection{Heavy-quark signatures}\label{sc:monoother}
%%%%%%%%%%%%%%%%%%%%%%%%%%%%%%%%%%%%%%%%%%%%%%%%%%%%%%%%%%%%%%%%%%%%%%%%%%%%%%%%%%%%%%%%%%%%%%%%%%%%%%%%%%%%

For operators generated by the exchange of a scalar mediator, couplings to light quarks are suppressed and the prospect of probing such interactions through the inclusive monojet channel at the LHC is limited. Dedicated searches focusing on bottom and top quark final states, occurring in processes as the ones shown in Fig.~\ref{fg:mono-tb}, have been proposed~\cite{mono-top1,mono-top2} to constrain this class of operators. A search in mono $b$-jets can significantly improve the current limits. The mono-$b$ signal arises partly from direct production of $b$-quarks in association with dark matter (Fig.~\ref{fg:mono-b}), but the dominant component is from top-quark pair production in the kinematic regime where one top is boosted (Fig.~\ref{fg:di-t}). A search for a top-quark pair + \met\ can strengthen the bounds even more; in this case signal and background would have very different missing energy distributions, providing a handle to disentangle one from the other.

\begin{figure}[ht]
  \centering
  \subfloat[$gb\to\x\bar{\x} + b$]{\label{fg:mono-b}\includegraphics[width=0.28\textwidth]{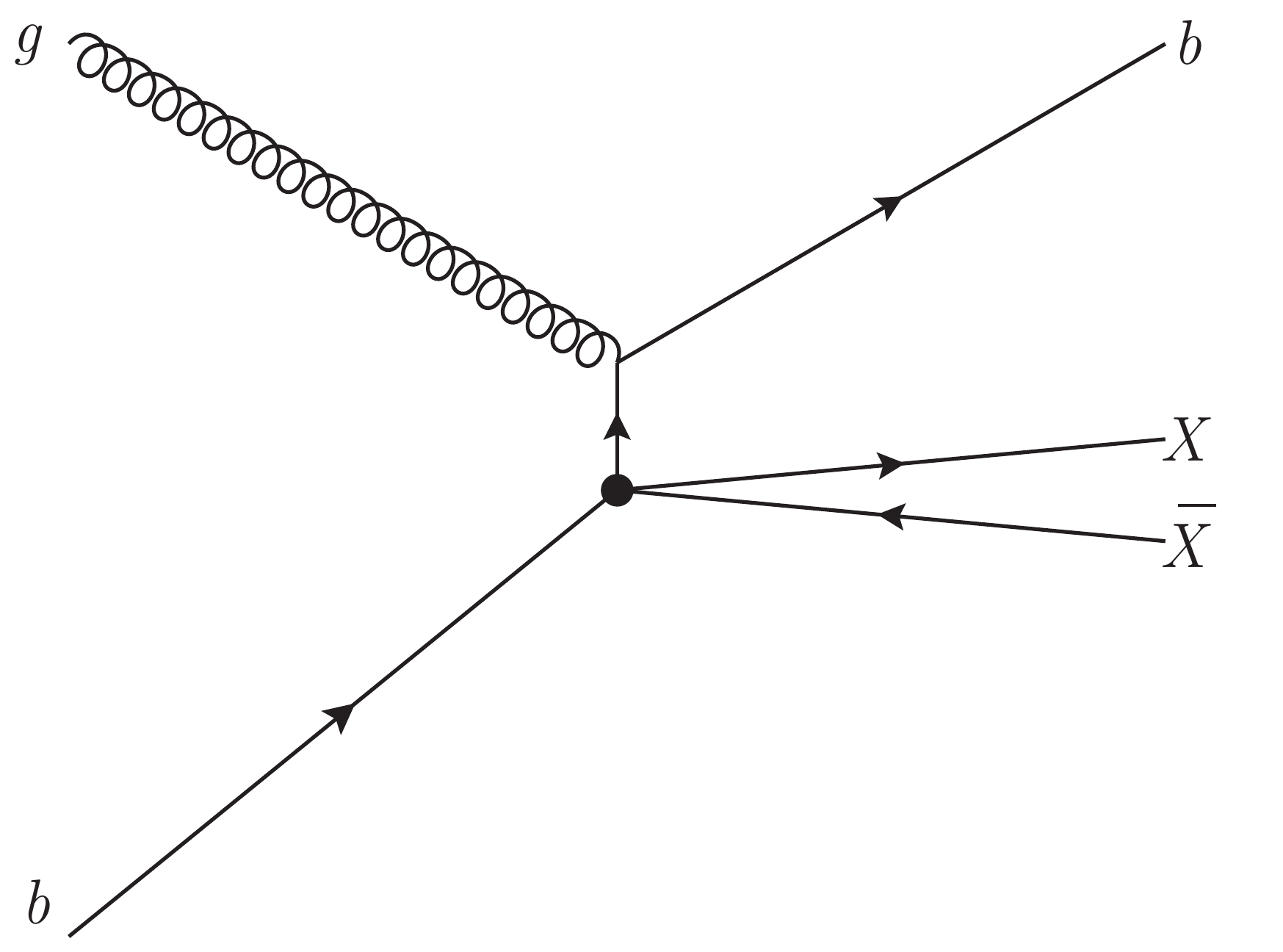}}\qquad
  \subfloat[$gg\to\x\bar{\x} + b\bar{b}$]{\label{fg:di-b}\includegraphics[width=0.28\textwidth]{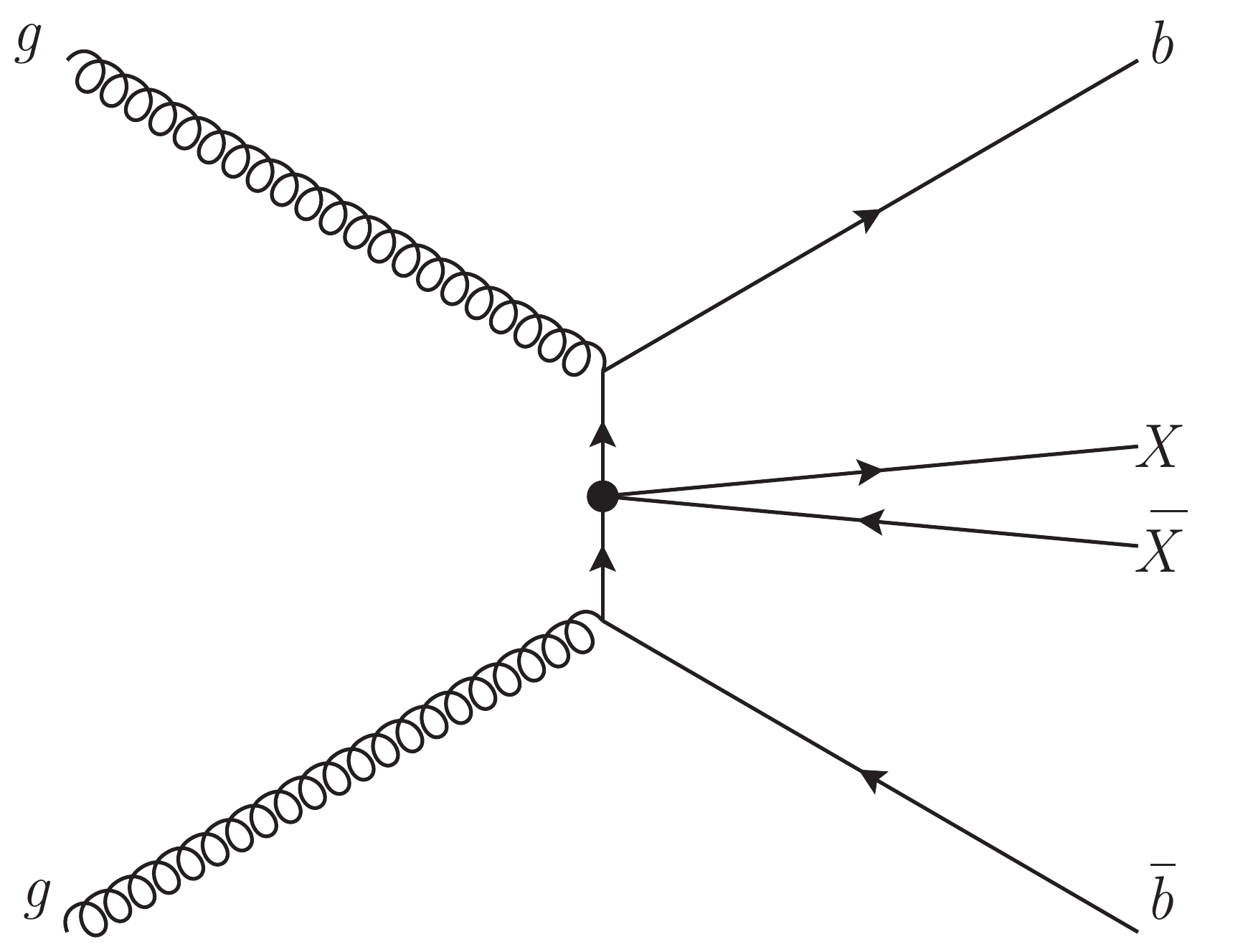}}\qquad
  \subfloat[$gg\to\x\bar{\x} + t\bar{t}$]{\label{fg:di-t}\includegraphics[width=0.28\textwidth]{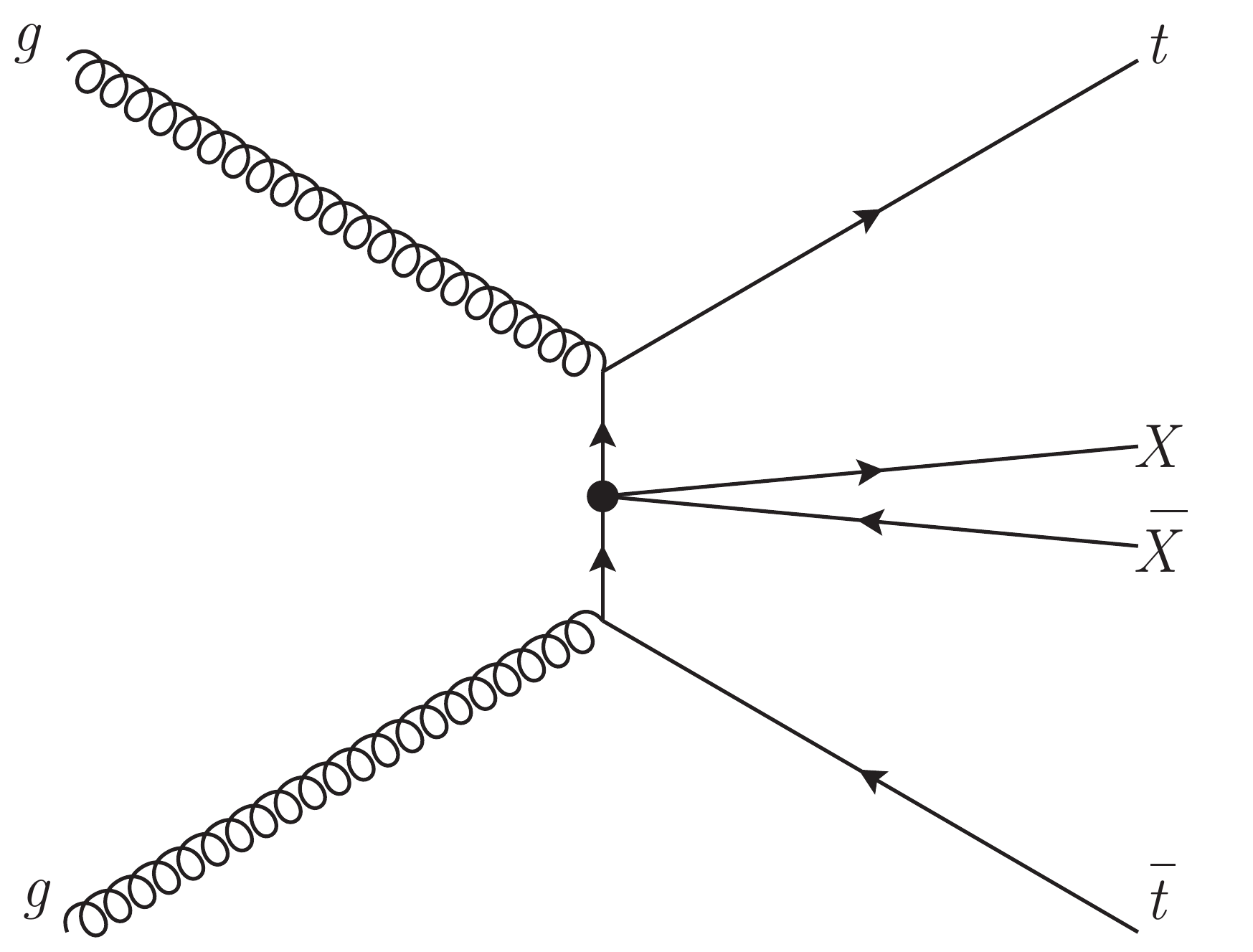}}\\ 
  \caption{Some of the dominant diagrams contributing to WIMP associated production with (a) a bottom, (b) two bottom quarks and (c) a top quark pair. From Ref.~\citen{mono-top1}.} \label{fg:mono-tb}
\end{figure}

%%%%%%%%%%%%%%%%%%%%%%%%%%%%%%%%%%%%%%%%%%%%%%%%%%%%%%%%%%%%%%%%%%%%%%%%%%%%%%%%%%%%%%%%%%%%%%%%%%%%%%%%%%%%
%%%%%%%%%%%%%%%%%%%%%%%%%%%%%%%%%%%%%%%%%%%%%%%%%%%%%%%%%%%%%%%%%%%%%%%%%%%%%%%%%%%%%%%%%%%%%%%%%%%%%%%%%%%%
\section{Searches for Supersymmetry}\label{sc:susy}
%%%%%%%%%%%%%%%%%%%%%%%%%%%%%%%%%%%%%%%%%%%%%%%%%%%%%%%%%%%%%%%%%%%%%%%%%%%%%%%%%%%%%%%%%%%%%%%%%%%%%%%%%%%%
%%%%%%%%%%%%%%%%%%%%%%%%%%%%%%%%%%%%%%%%%%%%%%%%%%%%%%%%%%%%%%%%%%%%%%%%%%%%%%%%%%%%%%%%%%%%%%%%%%%%%%%%%%%%

Supersymmetry (SUSY)~\cite{susy} is an extension of the Standard Model which assigns to each SM field a superpartner field with a spin differing by a half unit. SUSY provides elegant solutions to several open issues in the SM, such as the hierarchy problem and the grand unification. In particular, SUSY predicts the existence of a stable weakly interacting particle ---the lightest supersymmetric particle (LSP)--- that has the pertinent properties to be a dark matter particle, thus providing a very compelling argument in favor of SUSY.

SUSY searches in ATLAS~\cite{atlas-det} and CMS~\cite{cms-det} experiments typically focus on events with high transverse missing energy, which can arise from (weakly interacting) LSPs, in the case of \R-parity conserving SUSY, or from neutrinos produced in LSP decays, if \R-parity is broken (c.f.\ Section~\ref{sc:rpv}). Hence, the event selection criteria of inclusive channels are based on large \met, no or few leptons ($e$, $\mu$), many jets and/or $b$-jets, $\tau$-leptons and photons. In addition, kinematical variables such as the transverse mass, \mt, and the effective mass, \mef, assist in distinguishing further SUSY from SM events, whilst the \emph{effective transverse energy}~\cite{alberto-ete} can be useful to cross-check results, allowing a better and more robust identification of the SUSY mass scale, should a positive signal is found. Although the majority of the analysis simply look for an excess of events over the SM background, there is an increasing application of distribution shape fitting techniques~\cite{shape}.  

Typical SM backgrounds are top-quark production  ---including single-top---, $W$/$Z$ in association with jets, dibosons and QCD multijet events. These are estimated using semi- or fully data-driven techniques. Although the various analyses are optimized for a specific SUSY scenario, the interpretation of the results are extended to various SUSY models or topologies. 

Analyses exploring $R$-parity conserving SUSY models at LHC are roughly divided into inclusive searches for squarks and gluinos, for third-generation fermions, and for electroweak production (pairs of $\tilde{\chi}^0$, $\tilde{\chi}^{\pm}$, $\tilde{\ell}$). Although these searches are designed and optimised to look for $R$-parity conserving SUSY, interpretation in terms of $R$-parity violating (RPV) models is also possible. Other analyses are purely motivated by oriented by RPV scenarios and/or by the expectation of long-lived sparticles. Recent summary results from each category of ATLAS and CMS searches are presented in this section. 

%%%%%%%%%%%%%%%%%%%%%%%%%%%%%%%%%%%%%%%%%%%%%%%%%%%%%%%%%%%%%%%%%%%%%%%%%%%%%%%%%%%%%%%%%%%%%%%%%%%%%%%%%%%%
\subsection{Gluinos and first two generations of quarks}\label{sc:strong}
%%%%%%%%%%%%%%%%%%%%%%%%%%%%%%%%%%%%%%%%%%%%%%%%%%%%%%%%%%%%%%%%%%%%%%%%%%%%%%%%%%%%%%%%%%%%%%%%%%%%%%%%%%%%

At the LHC, supersymmetric particles are expected to be predominantly produced hadronically, i.e.\ through gluino-pair, squark-pair and squark-gluino production. Each of these (heavy) sparticles is going to decay into lighter ones in a cascade decay that finally leads to an LSP, which in most of the scenarios considered is the lightest neutralino \X. The two LSPs would escape detection giving rise to high transverse missing energy, hence the search strategy followed is based on the detection of high \met, many jets and possibly energetic leptons. The analyses make extensive use of data-driven Standard Model background measurements. 

The most powerful of the existing searches are based on all-hadronic final states with large missing transverse momentum~\cite{atlas-0l,cms-0l1,cms-0l2}. In the 0-lepton search, events are selected based on a jet+\met\ trigger, applying a lepton veto, requiring a minimum number of jets, high \met, and large azimuthal separation between the \met\ and reconstructed jets, in order to reject multijet background. In addition, searches for squark and gluino production in a final state with one or two leptons have been performed~\cite{atlas-lep,cms-lep1,cms-lep2}. The events are categorized by whether the leptons have higher or lower momentum and are referred to as the \emph{hard} and \emph{soft} lepton channels respectively. The soft-lepton analysis which enhances the sensitivity of the search in the difficult kinematic region where the neutralino and gluino masses are close to each other forming the so-called \emph{compressed spectrum.}~\cite{compressed1,compressed2} Leptons in the soft category are characterized by low lepton-\pt\ thresholds ($6-10~\gev$) and such events are triggered by sufficient \met. Hard leptons pass a threshold of $\sim25~\gev$ and are seeded with both lepton and \met\ triggers. Analyses based on the \emph{razor}~\cite{razor} variable have also been carried out by both experiments~\cite{atlas-razor,cms-razor}.

The major backgrounds ($t\bar{t}$, $W$+jets, $Z$+jets) are estimated by isolating each of them in a dedicated control region, normalizing the simulation to data in that control region, and then using the simulation to extrapolate the background expectations into the signal region. The multijet background is determined from the data by a matrix method. All other (smaller) backgrounds are estimated entirely from the simulation, using the most accurate theoretical cross sections available. To account for the cross-contamination of physics processes across control regions, the final estimate of the background is obtained with a simultaneous, combined fit to all control regions.

In the absence of deviations from SM predictions, limits for squark and gluino production are set. Figure~\ref{fg:msugra-gl-ttlsp} (left) illustrates the 95\% CL limits set by ATLAS under the minimal Supergravity (mSUGRA) model in the $(m_0,\,m_{1/2})$ plane with the 0-lepton plus \met\ plus multijets analysis~\cite{atlas-0l}. The remaining parameters are set to $\tan\beta = 30$, $A_0 = -2\,m_0$, $\mu > 0$, so as to acquire parameter-space points where the predicted mass of the lightest Higgs boson, $h_0$, is near $125~\gev$, i.e.\ compatible with the recently observed Higgs-like boson~\cite{higgs-disc1,higgs-disc2,higgs-prop1,higgs-prop2,higgs-prop3}. Exclusion limits are obtained by using the signal region with the best expected sensitivity at each point. By assumption, the mSUGRA model avoids both flavor-changing neutral currents and extra sources of $CP$ violation. For masses in the TeV range, it typically predicts too much cold dark matter, however these predictions depend of the presence of stringy effects that may dilute~\cite{mavro} or enhance~\cite{vergou1,vergou2} the predicted relic dark matter density. In the mSUGRA case, the limit on squark mass reaches 1750~\gev\ and on gluino mass is 1400~\gev\ if the results of various analyses are deployed~\cite{atlas-0l,atlas-msugra1,atlas-msugra2,atlas-msugra3,atlas-msugra4,atlas-msugra5,atlas-susy-results}. 

\begin{figure}[htb]
\centerline{\includegraphics[width=0.5\textwidth]{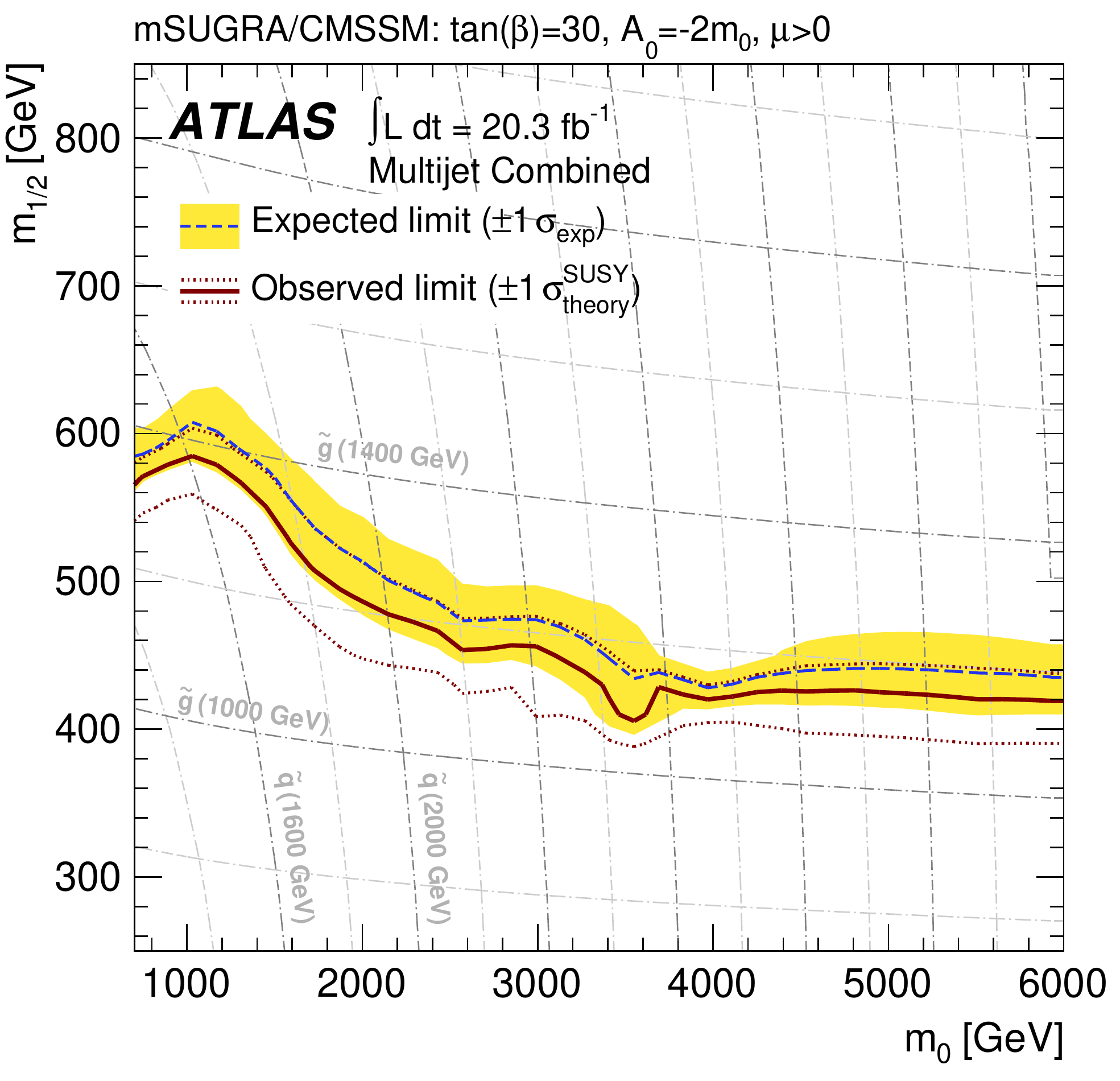}
  \hspace{0.0\textwidth}
  \includegraphics[width=0.5\textwidth]{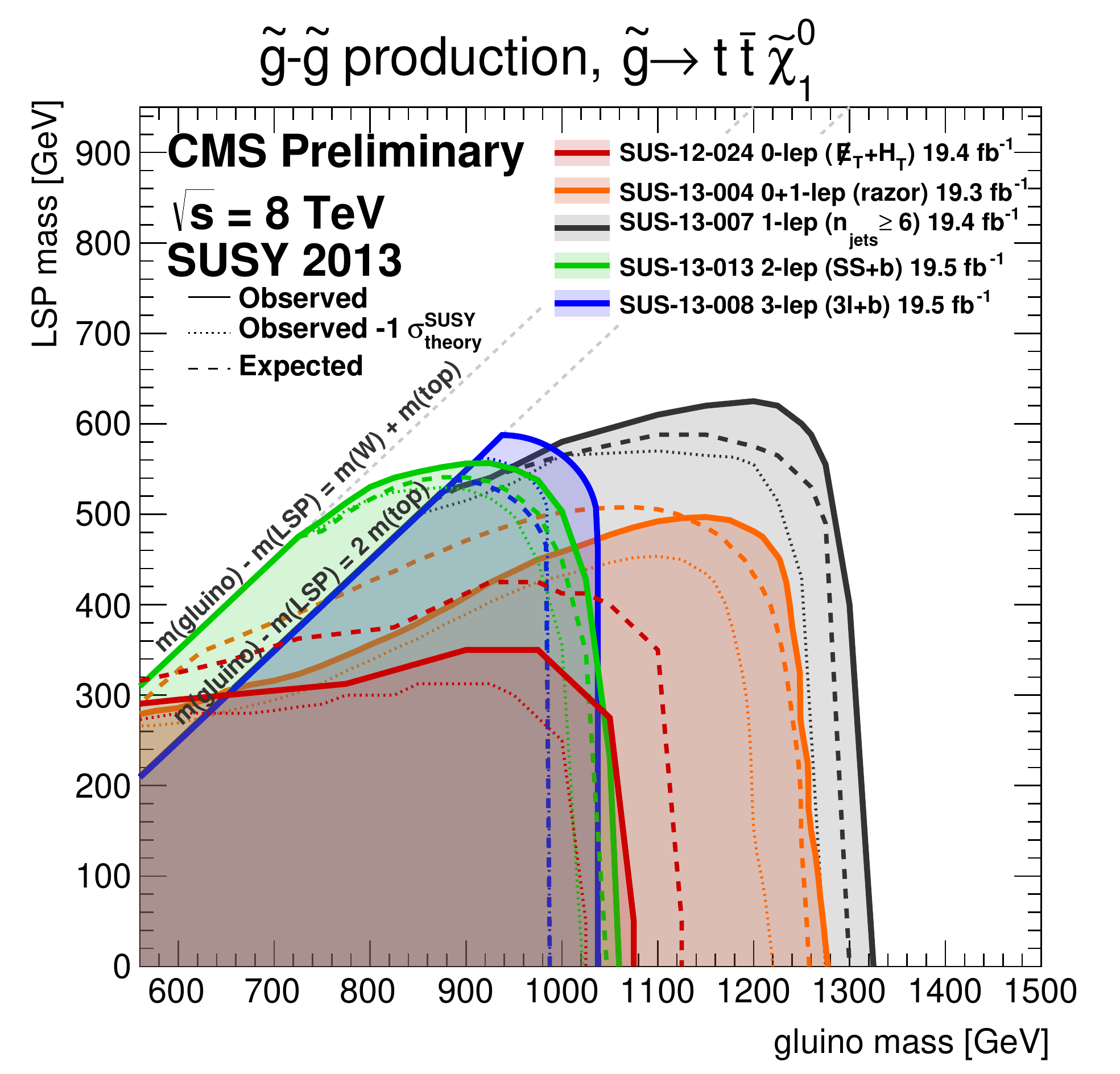}} 
\caption{\emph{Left:} Exclusion limits at 95\% CL for 8~\tev\ ATLAS multijets plus \met\ analysis in the $(m_0,\,m_{1/2})$ plane for the mSUGRA model. From Ref.~\citen{atlas-0l}. \emph{Right:} Summary of observed and expected limits~\cite{cms-razor,cms-gluino1,cms-gluino2,cms-gluino3,cms-ss2l} for gluino pair production with gluino decaying via a 3-body decay to a top, an anti-top and a neutralino. From Ref.~\citen{cms-susy-results}.\label{fg:msugra-gl-ttlsp}}
\end{figure}

%%%%%%%%%%%%%%%%%%%%%%%%%%%%%%%%%%%%%%%%%%%%%%%%%%%%%%%%%%%%%%%%%%%%%%%%%%%%%%%%%%%%%%%%%%%%%%%%%%%%%%%%%%%%
\subsection{Third-generation squarks}\label{sc:third}
%%%%%%%%%%%%%%%%%%%%%%%%%%%%%%%%%%%%%%%%%%%%%%%%%%%%%%%%%%%%%%%%%%%%%%%%%%%%%%%%%%%%%%%%%%%%%%%%%%%%%%%%%%%%

The previously presented limits from inclusive channels indicate that the masses of gluinos and first/second generation squarks are expected to be above 1~\tev. Nevertheless, in order to solve the hierarchy problem in a \emph{natural} way, the masses of the stops, sbottoms, higgsinos and gluinos need to be below the TeV-scale to properly cancel the divergences in the Higgs mass radiative corrections. Despite their production cross sections being smaller than for the first and second generation squarks, stop and sbottom may well be directly produced at the LHC and could provide the only direct observation of SUSY at the LHC in case the other sparticles are outside of the LHC energy reach. The lightest mass eigenstates of the sbottom and stop particles, $\t{b}_1$ and $\t{t}_1$, could hence be produced either directly in pairs or through $\t{g}$ pair production followed by $\t{g}\to\t{b}_1b$ or $\t{g}\to\t{t}_1t$ decays. Both cases will be discussed in the following.

For the aforementioned reasons, direct searches for third generation squarks have become a priority in both ATLAS and CMS. Such events are characterized by several energetic jets (some of them $b$-jets), possibly accompanied by light leptons, as well as high \met. A suite of channels have been considered, depending on the topologies allowed and the exclusions generally come with some assumptions driven by the shortcomings of the techniques and variables used, such as the requirement of 100\% branching ratios into particular decay modes.

In the case of the gluino-mediated production of stops, a simplified scenario (``Gtt model''), where $\tilde{t}_1$ is the lightest squark but $m_{\tilde{g}} < m_{\tilde{t}_1}$, has been considered. Pair production of gluinos is the only process taken into account since the mass of all other sparticles apart from the $\tilde{\chi}_1^0$ are above the \tev\ scale. A three-body decay via off-shell stop is assumed for the gluino, yielding a 100\% branching ratio for the decay $\tilde{g}\rightarrow t\bar{t}\tilde{\chi}_1^0$. The stop mass has no impact on the kinematics of the decay and the exclusion limits~\cite{cms-razor,cms-gluino1,cms-gluino2,cms-gluino3,cms-ss2l} set by the CMS experiment are presented in the $(m_{\tilde{g}},m_{\tilde{\chi}_1^0})$ plane in the right panel of Fig.~\ref{fg:msugra-gl-ttlsp}. For a massless LSP, gluinos with masses from 560~\gev\ to 1320~\gev\ are excluded.  

If the gluino is also too heavy to be produced at the LHC, the only remaining possibility is the direct $\t{t}_1\t{t}_1$ and $\t{b}_1\t{b}_1$ production. If stop pairs are considered, two decay channels can be distinguished depending on the mass of the stop: $\t{t}_{1}\to b\t{\chi}_1^{\pm}$ and $\t{t}_{1}\to t\t{\chi}_1^0$, as shown in the diagrams in Fig.~\ref{fg:directstop}. CMS and ATLAS carried out a wide range of different analyses in each of these modes at both 7~\tev\ and 8~\tev\ center-of-mass energy. In all these searches, the number of observed events has been found to be consistent with the SM expectation. Limits have been set by ATLAS on the mass of the scalar top for different assumptions on the mass hierarchy scalar top-chargino-lightest neutralino~\cite{atlas-tt1,atlas-tt2,atlas-tt3,atlas-tt4,atlas-tt5,atlas-tt6,atlas-tt7,atlas-tt8,atlas-tt9,atlas-tt10,atlas-tt11}. A scalar top quark of mass of up to 480~\gev\ is excluded at 95\% CL for a massless neutralino and a 150~\gev\ chargino. For a 300~\gev\ scalar top quark and a 290~\gev\ chargino, models with a neutralino with mass lower than 175~\gev\ are excluded at 95\% CL. 

\begin{figure}[htb]
\centerline{
\begin{minipage}[b]{0.4\linewidth}
\includegraphics[width=\textwidth]{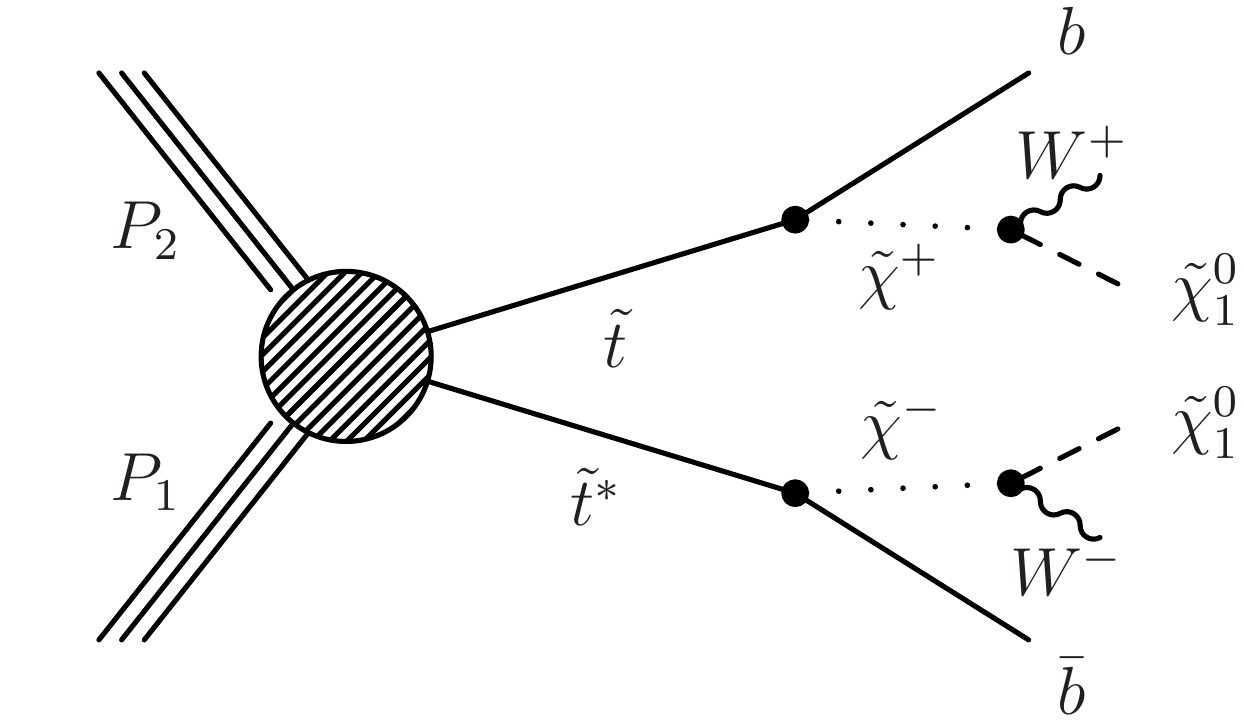} \vfill \includegraphics[width=\textwidth]{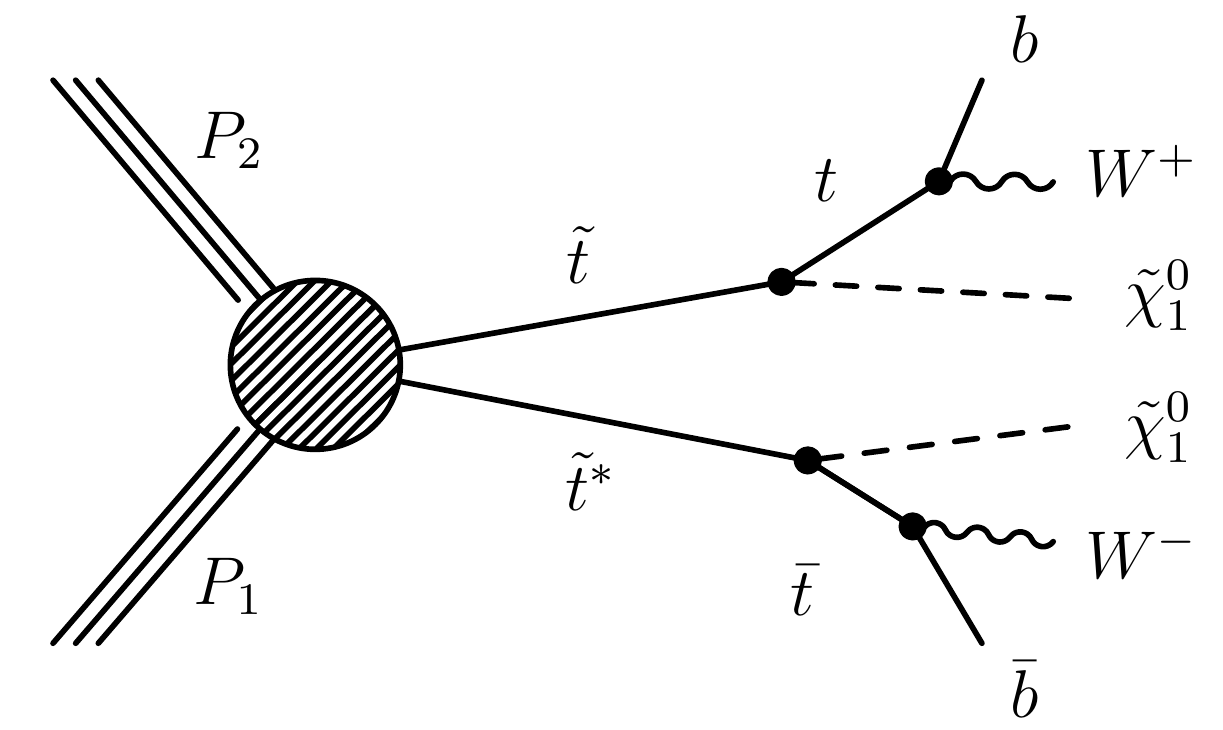}
 \end{minipage}
   \hspace{0.05\textwidth}\includegraphics[width=0.55\textwidth]{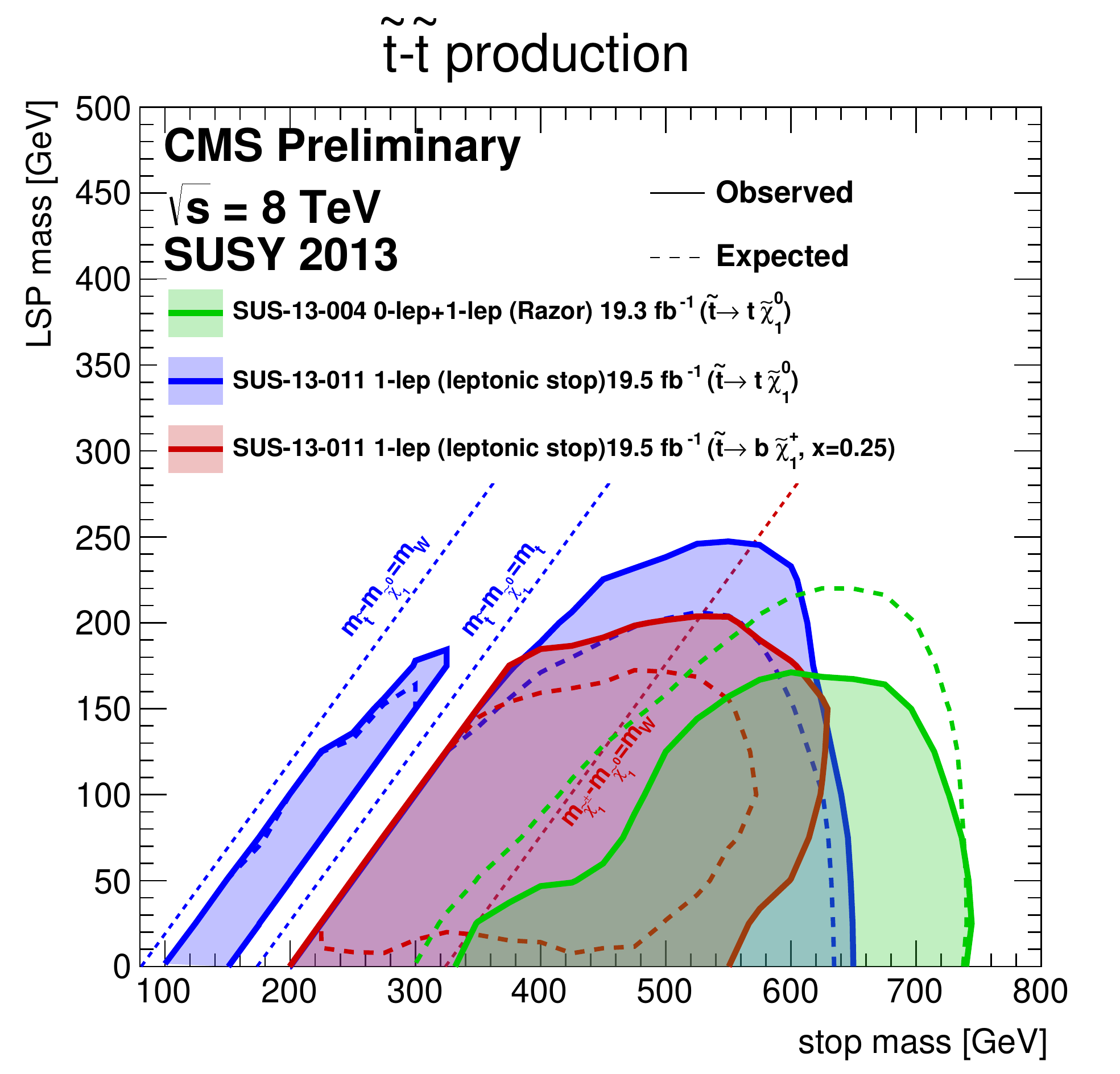} }
\caption{{\it Left:} Diagrams of $\t{t}_1\t{t}_1$ direct production with decays $\t{t}_{1}\to b\t{\chi}_1^{\pm}$ (top) and $\t{t}_{1}\to t\t{\chi}_1^0$ (bottom). {\it Right:} Summary of the dedicated CMS searches~\cite{cms-tt1,cms-tt2} for stop pair production based on $pp$ collision data taken at $\sqrt{s} = 8~\tev$. Exclusion limits at 95\% CL are shown in the $(\t{t}_{1},\,\t{\chi}_1^0)$ mass plane for channels targeting $\t{t}_{1}\to b\t{\chi}_1^{\pm}$ and $\t{t}_{1}\to t\t{\chi}_1^0$. The dashed and solid lines show the expected and observed limits, respectively. From Ref.~\citen{cms-susy-results}.\label{fg:directstop}}
\end{figure}

For the case of a high-mass stop decaying to a top and neutralino ($\t{t}_{1}\to t\t{\chi}_1^0$), analyses requiring one, two or three isolated leptons, jets and large \met\ have been carried out. No significant excess of events above the rate predicted by the SM is observed and 95\% CL upper limits are set on the stop mass in the stop-neutralino mass plane. The region of excluded stop and neutralino masses is shown on the right panel of Fig.~\ref{fg:directstop} for the CMS analyses~\cite{cms-tt1,cms-tt2}. Stop masses are excluded between 200~\gev\ and 750~\gev\ for massless neutralinos, and stop masses around 500~\gev\ are excluded along a line which approximately corresponds to neutralino masses up to 250~\gev. 

%%%%%%%%%%%%%%%%%%%%%%%%%%%%%%%%%%%%%%%%%%%%%%%%%%%%%%%%%%%%%%%%%%%%%%%%%%%%%%%%%%%%%%%%%%%%%%%%%%%%%%%%%%%%
\subsection{Electroweak gaugino production}\label{sc:gaugino}
%%%%%%%%%%%%%%%%%%%%%%%%%%%%%%%%%%%%%%%%%%%%%%%%%%%%%%%%%%%%%%%%%%%%%%%%%%%%%%%%%%%%%%%%%%%%%%%%%%%%%%%%%%%%

If all squarks and gluinos are above the TeV scale, weak gauginos with masses of few hundred GeV may be the only sparticles accessible at the LHC. As an example, at $\sqrt{s}Ê= 7~\tev$, the cross-section of the associated production $\t{\chi}_1^{\pm}\t{\chi}_2^0$ with degenerate masses of 200~\gev\ is above the 1-\tev\ gluino-gluino production cross section by one order of magnitude. Chargino pair production is searched for in events with two opposite-sign leptons and \met\ using a jet veto, through the decay $\t{\chi}_1^{\pm} \to \ell^{\pm}\nu\t{\chi}_1^0$. A summary of related analyses~\cite{cms-ewkino1,cms-ewkino2} performed by CMS is shown in Fig.~\ref{fg:ewkino}. Charginos with masses between 140 and 560~\gev\ are excluded for a massless LSP in the chargino-pair production with an intermediate slepton/sneutrino between the $\t{\chi}_1^{\pm}$ and the $\t{\chi}_1^0$. If $\t{\chi}_1^{\pm}\t{\chi}_2^0$ production is assumed instead, the limits range from 11 to 760~\gev. The corresponding limits involving intermediate $W$, $Z$ and/or $H$ are significantly weaker.  

\begin{figure}[htb]
\centerline{\includegraphics[width=0.65\textwidth]{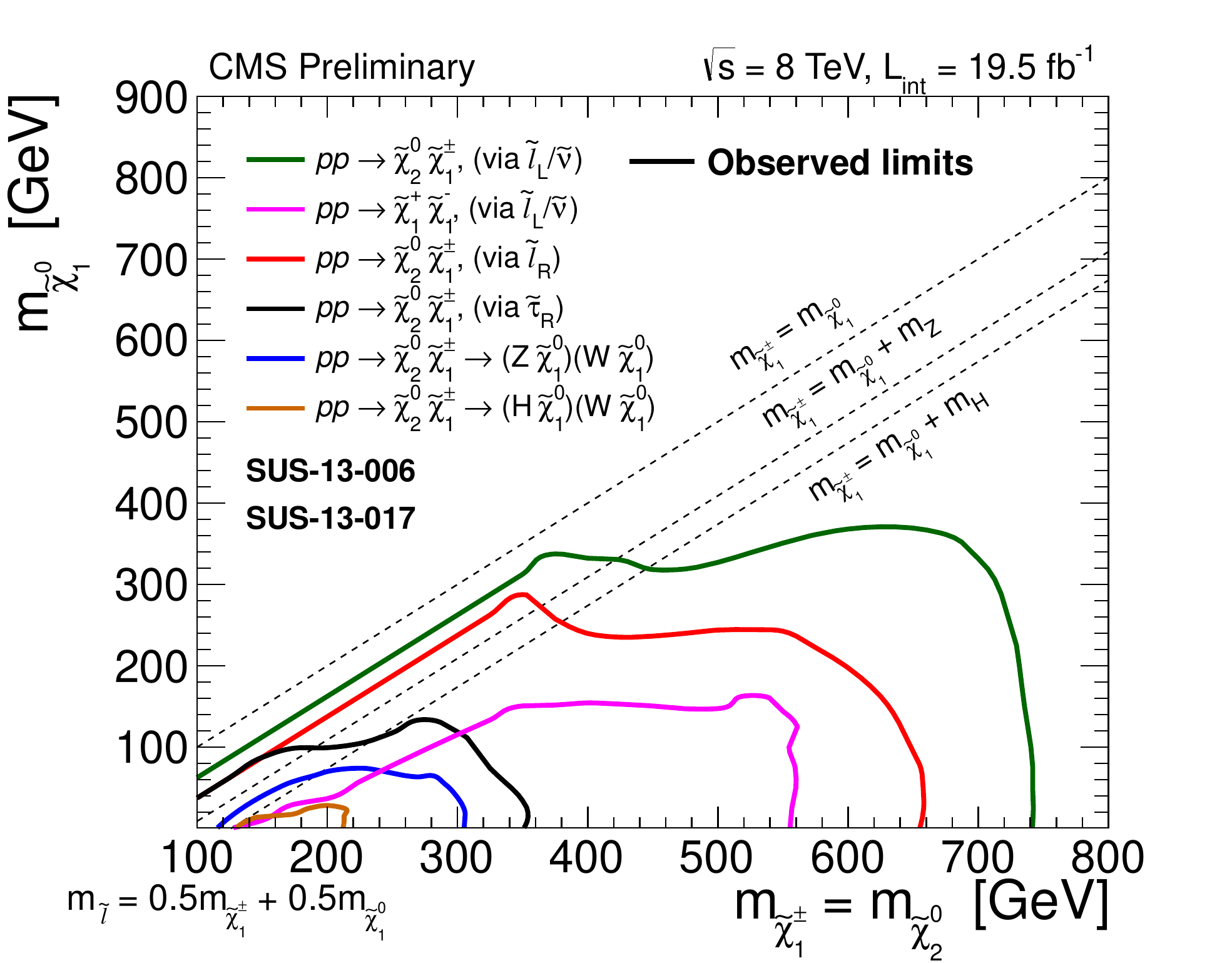}}
\caption{Summary of observed limits for electroweak-gaugino production from CMS~\cite{cms-ewkino1,cms-ewkino2}. From Ref.~\citen{cms-susy-results}.\label{fg:ewkino}}
\end{figure}

In several analyses the EW sector of the MSSM has been studied for parameter choices that yield the correct DM relic density. In Ref.~\citen{ewk-1}, the constraints coming from the trilepton/dilepton search by ATLAS and CMS from direct pair production of chargino and neutralino or slepton pair production have been considered and the implication on DM and collider searches have been investigated, while in Ref.~\citen{ewk-2} we have examined the search prospects of DM-allowed SUSY signals with several models in the light of LHC data. 

%%%%%%%%%%%%%%%%%%%%%%%%%%%%%%%%%%%%%%%%%%%%%%%%%%%%%%%%%%%%%%%%%%%%%%%%%%%%%%%%%%%%%%%%%%%%%%%%%%%%%%%%%%%%
\subsection{$R$-parity violating SUSY and meta-stable sparticles}\label{sc:rpv}
%%%%%%%%%%%%%%%%%%%%%%%%%%%%%%%%%%%%%%%%%%%%%%%%%%%%%%%%%%%%%%%%%%%%%%%%%%%%%%%%%%%%%%%%%%%%%%%%%%%%%%%%%%%%

\R-parity is defined as: $R = (-1)^{3(B-L)+2S}$, where $B$, $L$ and $S$ are the baryon number, lepton number and spin, respectively. Hence $R=+1$ for all Standard Model particles and $R=-1$ for all SUSY particles. It is stressed that the conservation of \R-parity is an \emph{ad-hoc} assumption. The only firm restriction comes from the proton lifetime: non-conservation of both $B$ and $L$ leads to rapid proton decay. \R-parity conservation has serious consequences in SUSY phenomenology in colliders: the SUSY particles are produced in pairs and the lightest SUSY particle is absolutely stable, thus providing a WIMP candidate. Here we highlight the status of RPV supersymmetry~\cite{rpv} searches at the LHC.

Both ATLAS and CMS experiments have probed RPV SUSY through various channels, either by exclusively searching for specific decay chains, or by inclusively searching for multilepton events. ATLAS has looked for resonant production of $e\mu$, $e\tau$ and $\mu\tau$~\cite{atlas-rpv-emu1,atlas-rpv-emu2,atlas-rpv-emu3}, for multijets~\cite{atlas-rpv-multijets}, for events with at least four leptons~\cite{atlas-rpv-4l} and for excesses in the $e\mu$ continuum~\cite{atlas-rpv-emu-cont}. Null inclusive searches in the one-lepton channel~\cite{atlas-brpv1,atlas-brpv2} have also been interpreted in the context of a model where RPV is induced through bilinear terms~\cite{brpv1,brpv2,brpv3,brpv4,brpv5}.

Recent CMS analyses are focused on studying the lepton number violating terms $\lambda_{ijk}L_iL_j\bar{e}_{k}$ and $\lambda'_{ijk}L_iQ_j\bar{d}_{k}$, which result in specific signatures involving leptons in events produced in $pp$ collisions at LHC. A search for resonant production and the following decay of $\t{\mu}$ which is caused by $\lambda'_{211}\neq0$ has been conducted~\cite{cms-rpv-ssmu}. Multilepton signatures caused by LSP decays due to various $\lambda$ and $\lambda'$ terms in stop production have been probed~\cite{cms-rpv-stop}. Ref.~\citen{cms-rpv-4l} discusses the possibility of the generic model independent search for RPV SUSY in 4-lepton events. A summary of the limits set by several CMS analyses~\cite{cms-ss2l,cms-rpv-stop,cms-rpv1,cms-rpv2,cms-rpv3} are listed in Fig.~\ref{fg:rpv}.

\begin{figure}[htb]
\centerline{\includegraphics[width=\textwidth]{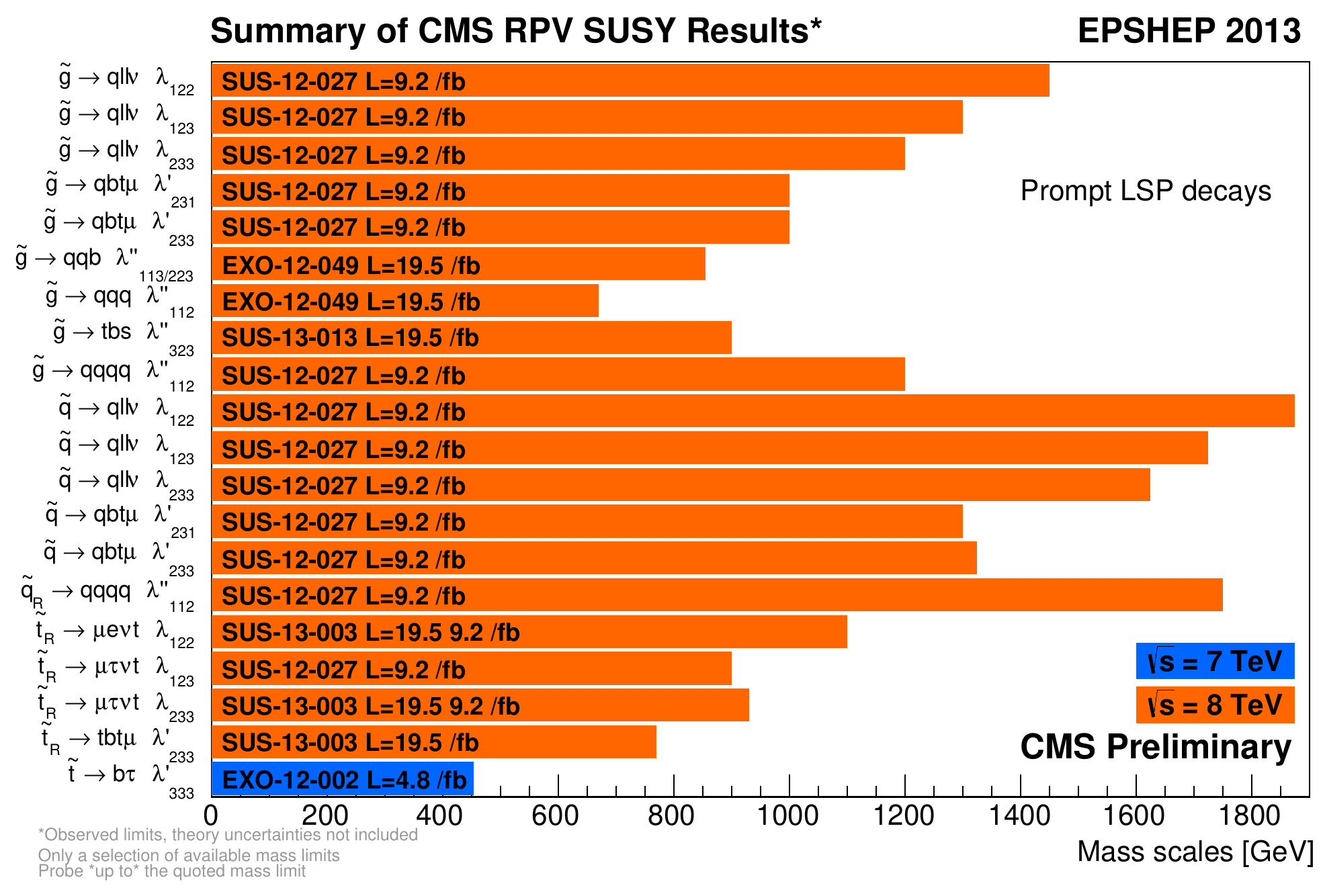}}
\caption{Best exclusion limits for the masses of the mother particles, for RPV scenarios, for each topology, for all CMS results~\cite{cms-ss2l,cms-rpv-stop,cms-rpv1,cms-rpv2,cms-rpv3}. In this plot, the lowest mass range is $m_{\text{mother}}=0$, but results are available starting from a certain mass depending on the analyses and topologies. Branching ratios of 100\% are assumed, values shown in plot are to be interpreted as upper bounds on the mass limits. From Ref.~\citen{cms-susy-results}.\label{fg:rpv}}
\end{figure}

In view of the null results in other SUSY searches, it became mandatory to fully explore the SUSY scenario predicting meta-stable or long-lived particles. These particles, not present in the Standard Model, would provide striking signatures in the detector and rely heavily on a detailed understanding of its performance. In SUSY, non-prompt particle decay can be caused by (i) very weak RPV~\cite{atlas-dv}, (ii) low mass difference between a SUSY particle and the LSP~\cite{atlas-kinked}, or (iii) very weak coupling to the gravitino in GMSB models~\cite{atlas-rhadrons,cms-rhadrons,atlas-nonp-phot,cms-nonp-phot}. A small part of these possibilities have been explored by the ATLAS~\cite{atlas-susy-results} and CMS~\cite{cms-susy-results} experiments covering specific cases, difficult to summarize here. There is still a wide panorama of signatures to be explored, in view of various proposed SUSY scenarios pointing towards this direction. 

As a last remark, we address the issue of (not necessarily cold) dark matter in RPV SUSY models. These seemingly incompatible concepts \emph{can} be reconciled in models with a gravitino~\cite{rpv-grav1,rpv-grav2,rpv-grav3} or an axino~\cite{rpv-axino} LSP with a lifetime exceeding the age of the Universe. In both cases, RPV is induced by bilinear terms in the superpotential that can also explain current data on neutrino masses and mixings without invoking any GUT-scale physics~\cite{brpv1,brpv2,brpv3,brpv4,brpv5}. Decays of the next-to-lightest superparticle occur rapidly via RPV interaction, and thus they do not upset the Big-Bang nucleosynthesis, unlike the \R-parity conserving case. Such gravitino DM is proposed in the context of $\mu\nu$SSM~\cite{munussm1,munussm2,munussm3} with profound prospects for detecting $\gamma$ rays from their decay~\cite{munussm-dm}. 

%%%%%%%%%%%%%%%%%%%%%%%%%%%%%%%%%%%%%%%%%%%%%%%%%%%%%%%%%%%%%%%%%%%%%%%%%%%%%%%%%%%%%%%%%%%%%%%%%%%%%%%%%%%%
%%%%%%%%%%%%%%%%%%%%%%%%%%%%%%%%%%%%%%%%%%%%%%%%%%%%%%%%%%%%%%%%%%%%%%%%%%%%%%%%%%%%%%%%%%%%%%%%%%%%%%%%%%%%
\section{Looking for Extra Dimensions}\label{sc:ed}
%%%%%%%%%%%%%%%%%%%%%%%%%%%%%%%%%%%%%%%%%%%%%%%%%%%%%%%%%%%%%%%%%%%%%%%%%%%%%%%%%%%%%%%%%%%%%%%%%%%%%%%%%%%%
%%%%%%%%%%%%%%%%%%%%%%%%%%%%%%%%%%%%%%%%%%%%%%%%%%%%%%%%%%%%%%%%%%%%%%%%%%%%%%%%%%%%%%%%%%%%%%%%%%%%%%%%%%%%

Theories with universal extra dimensions (UED)~\cite{ued} are very promising for solving shortcomings of the Standard Model, such as explaining the three fermion generations in terms of anomaly cancellations and providing a mechanism for an efficient suppression of the proton decay. In the UED framework, unlike in other proposed extra-dimensional models, all SM particles are postulated to propagate in a $\tev^{-1}$-sized \emph{bulk}, i.e.\ normal space plus the extra compactified dimensions. In addition, UED models can naturally incorporate a $Z_2$ symmetry called KK~parity, analogous to $R$~parity in supersymmetry, leading to a well-motivated dark matter\footnote{Other species of extra dimension models have been probed thoroughly with LHC data, however they do not provide a viable DM candidate, hence they are beyond the scope of this article.} candidate, the lightest KK particle~\cite{ued-dm}.

Indirect constraints on the compactification radius $R$ from electroweak precision tests and the dark matter relic density favor a mass scale for the first KK modes of ${\mathcal O}(1~\tev)$. Therefore UED models can be directly probed at the LHC, either through \met-based signatures or via searches for resonances near the \tev\ scale. Since the mass scale of the KK resonances is rather compressed, UED is only accessible through analyses based on soft leptons/jets and moderately-high missing transverse momentum~\cite{atlas-ued1,atlas-ued2}. 

The rich LHC phenomenology of UED models has been exploited to study the discovery reach or set limits based on already performed searches in leptonic~\cite{ued-lepton1,ued-lepton2,ued-lepton3,ued-lepton4,ued-flacke} final states, photon~\cite{ued-photon} channels and through the Higgs sector~\cite{ued-higgs,ued-belanger}. In particular, several limits have been set on the minimal UED model (mUED)~\cite{mued}, in which only the 5D~extensions of the SM operators are present at the cutoff scale $\Lambda$, whereas boundary operators and other higher-dimensional bulk operators are assumed to vanish at $\Lambda$. Existing CMS limits on the ratio ${\mathcal R}$, defined as
\begin{equation}\label{eq:ued}
{\mathcal R} = \frac{\sigma(pp\to Z'+X\to\ell\ell+X)}{\sigma(pp\to Z+X\to\ell\ell+X)},
\end{equation}
obtained by searching for resonances in the dilepton spectrum~\cite{cms-dilepton}, have been re-interpreted~\cite{ued-flacke} to set bounds on the mass of the $A^{(2)}$ mode, as shown in Fig.~\ref{fg:ued} (left). This way, lower limits on $m_{A^{(2)}}$ have been set at $\sim1400~\gev$. 

\begin{figure}[htb]
\centerline{\includegraphics[width=0.52\textwidth]{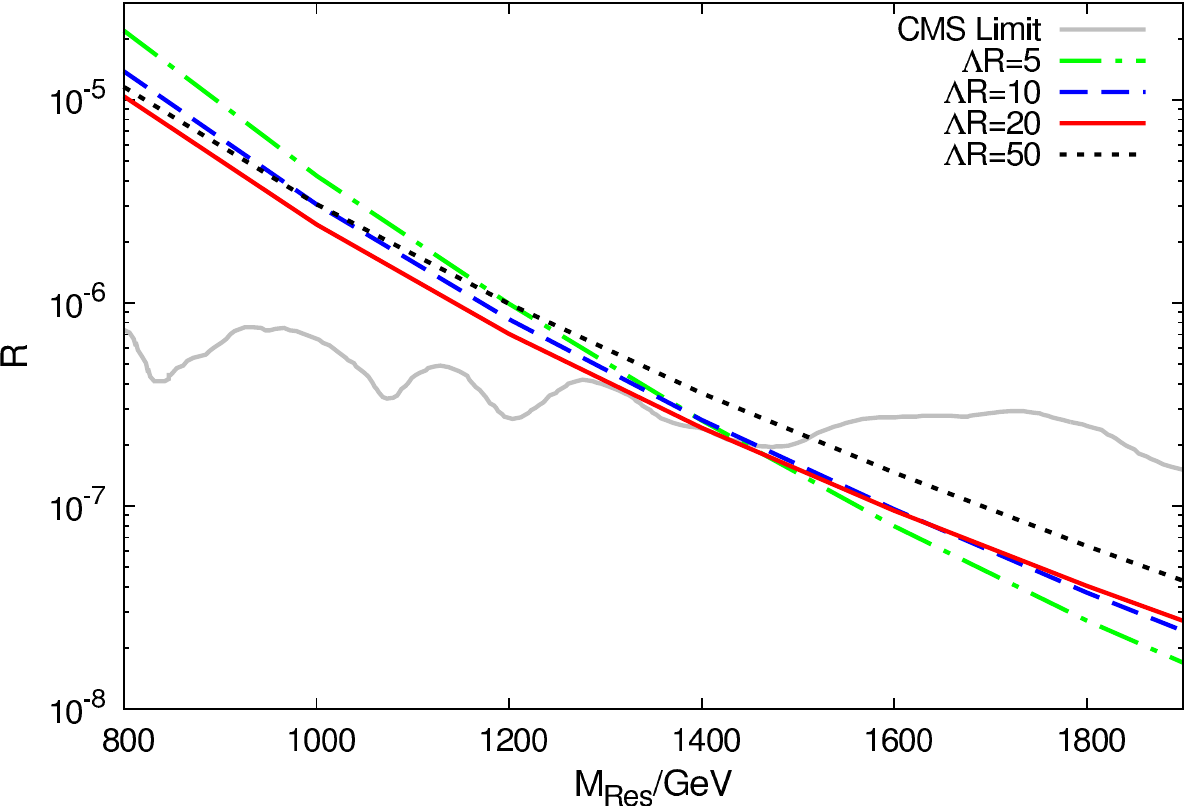}
  \hspace{0.01\textwidth}
  \includegraphics[width=0.47\textwidth]{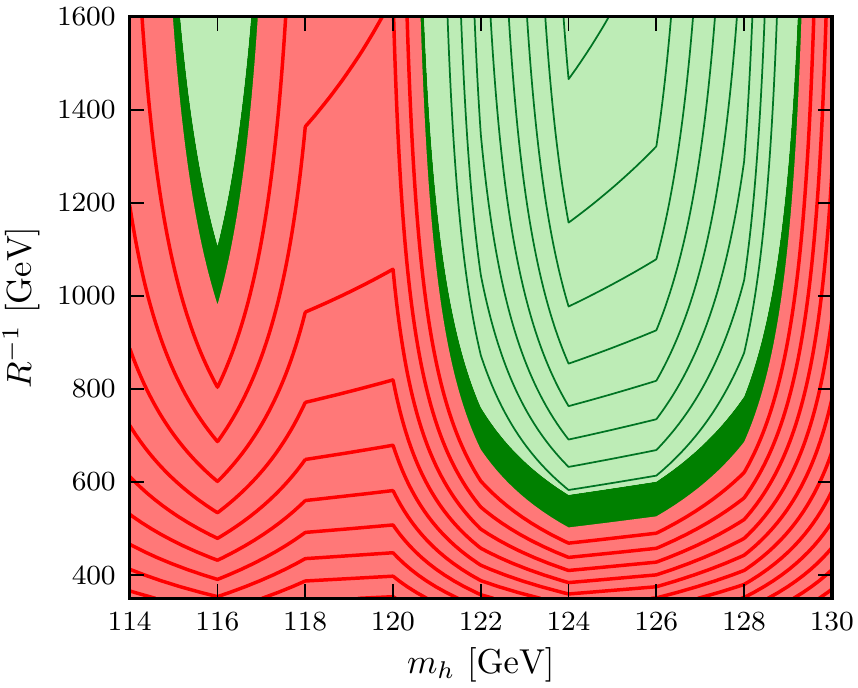}} 
\caption{{\it Left:} Ratio ${\mathcal R}$ defined in Eq.~(\ref{eq:ued}) for different benchmark points of the mUED model, as a function of the resonance mass $M_{\rm Res}\equiv m_{A^{(2)}}$. From Ref.~\citen{ued-flacke}. {\it Right:} 95\% CL exclusion limits in the mUED $(m_h,\,R^{-1})$ plane from Higgs boson searches at the LHC. The allowed region is in light green (light gray) and the excluded region is in red (medium gray). From Ref.~\citen{ued-belanger}.\label{fg:ued}}
\end{figure}

In another analysis~\cite{ued-belanger}, the Higgs sector of mUED is exploited to test this model at the LHC, 
by using combined ATLAS and CMS limits in the $gg\to h\to\gamma\gamma$, $gg\to h\to W^+W^-\to\ell^+\bar{\nu}\ell^-\nu$ and $gg\to h\to ZZ\to\ell^+\ell^-\ell^+\ell^-$ channels, based on 7~\tev\ and 8~\tev\ data. These limits lead to bounds on the mUED model in the $(m_h,\,R^{-1})$ plane, $m_h$ being the Higgs mass, as shown in Fig.~\ref{fg:ued} (right). It is found that $R^{-1}<550~\gev$ is excluded at 95\% CL, while for larger $R^{-1}$ only a very narrow ($\pm\,1-4~\gev$) mass window around $m_h=125~\gev$, i.e.\ the mass of the recently observed boson~\cite{higgs-disc1,higgs-disc2}, and another short window $\sim118~\gev$ (for $R^{-1}>1~\tev$) remain unconstrained.  

%%%%%%%%%%%%%%%%%%%%%%%%%%%%%%%%%%%%%%%%%%%%%%%%%%%%%%%%%%%%%%%%%%%%%%%%%%%%%%%%%%%%%%%%%%%%%%%%%%%%%%%%%%%%
\section{The Future: $e^+e^-$ Colliders}\label{sc:ilc}
%%%%%%%%%%%%%%%%%%%%%%%%%%%%%%%%%%%%%%%%%%%%%%%%%%%%%%%%%%%%%%%%%%%%%%%%%%%%%%%%%%%%%%%%%%%%%%%%%%%%%%%%%%%%

Linear $e^+e^-$ accelerators of the next generation, namely the ILC~\cite{ilc-phys-tdr} and the CLIC~\cite{clic-phys}, may have enough energy to produce and study WIMPs. The International Linear Collider (ILC)~\cite{ilc} is a $200-500~\gev$ ---extendable to 1~\tev--- center-of-mass high-luminosity linear $e^+e^-$ collider, based on $1.3~{\rm GHz}$ superconducting radio-frequency accelerating technology. The Compact Linear Collider (CLIC), on the other hand, is a \tev-scale high-luminosity linear electron-positron collider based on a novel two-beam technique providing acceleration gradients at the level of $100~{\rm MV/m}$. 

Positron-electron colliders can play a major role in providing precision data for understanding dark matter, should it be discovered in colliders, among other measurements, due to three characteristics: (i) all energy of incoming particles is transferred to the final-state particles, allowing the setting of severe constraints on the mass of invisible particles; (ii) the cross sections of all production processes are of the same order of magnitude, thus making the decays of the BSM particles clearly visible; and (iii) the energy, projectile and polarization of the beam can be tuned to choose the optimal configuration for the physics of interest. All these features are instrumental in pinning down the properties of DM in such colliders.

The study of model-independent production of WIMP pairs at the linear collider through the monophoton channel, $e^+e^-\to\x\bar{\x}\gamma$, has shown that a WIMP in the mass range of $\sim60-200~\gev$ can be discovered with a $5\sigma$ significance for an annihilation fraction of unity~\cite{dm-ilc}. In terms of the effective dark matter model, it is found that the ILC should be able to probe couplings of $10^{-7}~\gev^{-2}$ or $10^{-4}~\gev^{-1}$, depending on the mass dimension of the theory~\cite{ilc-illuminati}. In model predicting vector dark matter, the ILC may be able to probe even weaker couplings in the case of low DM mass. 

Once DM is detected through a non-gravitational interaction, the new-particle mass may be constrained through methods based on matching specific decay chains to measurements of kinematic edges in invariant-mass distributions of two or three reconstructed objects~\cite{ilc-mass,dm-colliders}. This is one way to overcome the unconstrained kinematics of the production of two invisible particles in conjunction with the measurement of the momentum spectrum of the final-state leptons and the scanning of the particle pair production thresholds~\cite{clic-susy-ued}. 

The determination of the spin of the new particle will play a major role in the identification of the DM nature. This issue has been studied thoroughly in the case of SUSY versus UED~\cite{clic-susy-ued}. Both models feature a stable particle that is a viable DM candidate: the lightest neutralino, $\t{\chi}_1^0$, in SUSY, and the lightest KK excitation of the photon, $\gamma^{(1)}$ in UED. The fact that similar decay chains lead to those WIMP candidates, while their spins are different, can be exploited to distinguish them. In that case, the difference in the distribution shape of the muon polar angle, $\theta_{\mu}$, for $e^+e^-\to\t{\mu}^+\t{\mu}^-\to\mu^+\mu^-\t{\chi}_1^0\t{\chi}_1^0$ and $e^+e^-\to\mu^{(1)}\mu^{(1)}\to\mu^+\mu^-\gamma^{(1)}\gamma^{(1)}$ is shown in Fig.~\ref{fg:lc} (left) for a study carried out for CLIC. 

\begin{figure}[htb]
\centerline{\includegraphics[width=0.5\textwidth]{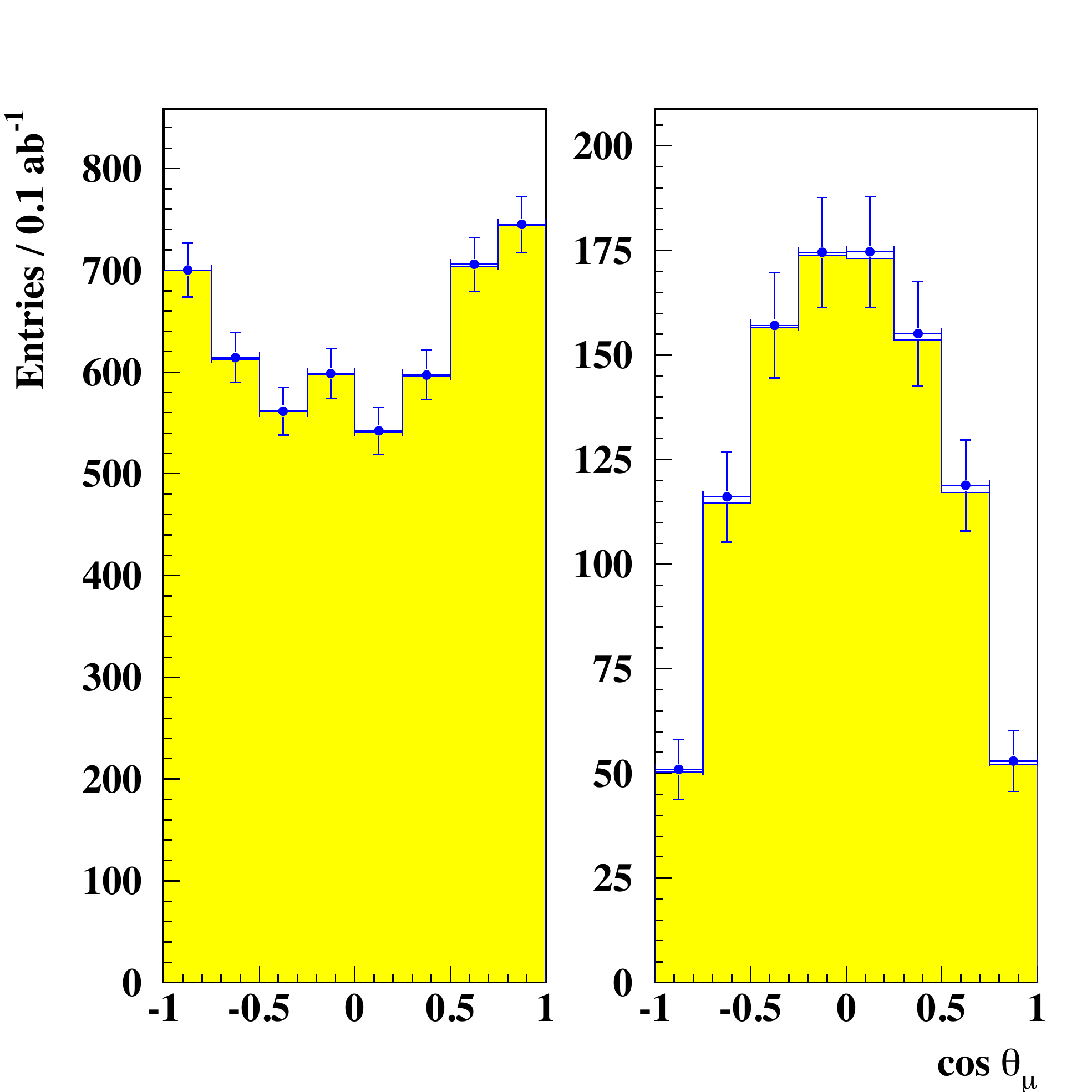}
  \hspace{0.0\textwidth}
  \includegraphics[width=0.5\textwidth]{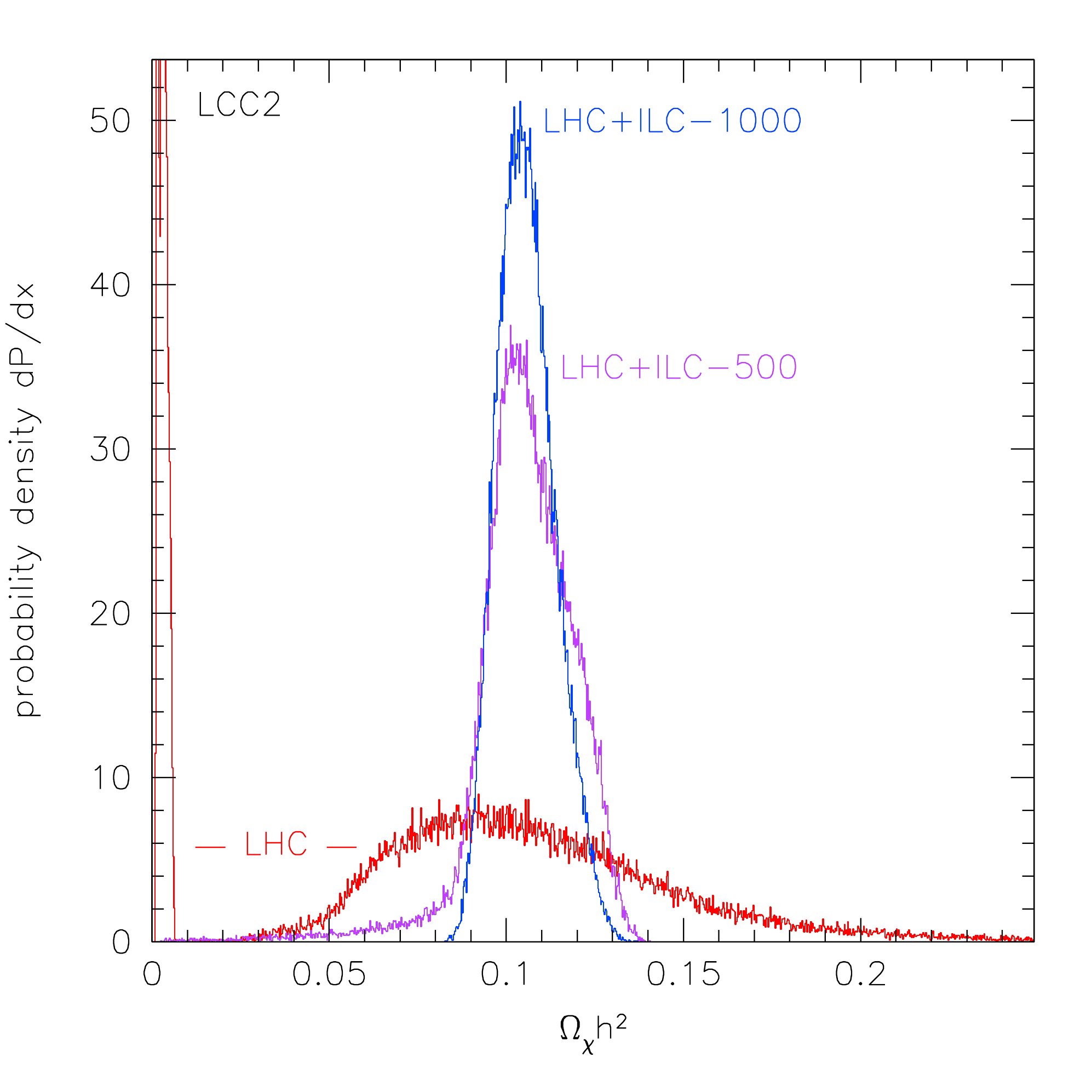}} 
\caption{{\it Left:} Differential cross section ${\rm d}\sigma/{\rm d}\!\cos\theta_{\mu}$ for UED (left) and supersymmetry (right) as a function of the muon scattering angle $\theta_{\mu}$, including the effects of  event selection, beamstrahlung and detector resolution and acceptance. The data points represent the sum of background and signal events, while the yellow (light grey) shaded area is the signal only. From Ref.~\citen{clic-susy-ued}. {\it Right:} Relic density for benchmark point LCC2. The three curves show the results for expected measurements from LHC (red), up-to-500~\gev\ ILC (magenta) and up-to-1~\tev\ ILC (blue). There are two overlapping very high peaks at $\Omega_{\x}h^2 < 0.01$, due to the wino and higgsino solutions to the LHC constraints. From Ref.~\citen{dm-colliders}.\label{fg:lc}}
\end{figure}

Having identified the nature of the underlying physics of the observed DM particle, an $e^+e^-$ collider can measure the mass and couplings of pair-produced particles and, in turn, to determine properties relevant to astroparticle physics, such as the WIMP relic density $\Omega_{\x}h^2$. In particular, the study in Ref.~\citen{dm-colliders} considers several mSUGRA benchmark points, representative of the variety of neutralino annihilation mechanisms, and by scanning the SUSY parameters determines the probability distribution function for the neutralino relic density, given various sparticle mass and yield measurements. This density for the point LCC2\footnote{LCC2 is a mSUGRA benchmark point with the following parameters and WIMP relic density: $m_0=3280~\gev$, $m_{1/2}=300~\gev$, $\tan\beta=10$, $\mu>0$, $A_0=0$, $\Omega_{\x}h^2=0.109$.} and for three different collider options (LHC, ILC500 and ILC1000) is shown in Fig.~\ref{fg:lc} (right). The distribution from the LHC constraints is quite broad, with a standard deviation of about 40\% and also a significant secondary peak near $\Omega_{\x}h^2\simeq0$. The prediction of $\Omega_{\x}h^2$ from the ILC data at 500~\gev\ has an accuracy of about 14\%, and this improves to about 8\% using the data from the ILC at 1000~\gev.

%%%%%%%%%%%%%%%%%%%%%%%%%%%%%%%%%%%%%%%%%%%%%%%%%%%%%%%%%%%%%%%%%%%%%%%%%%%%%%%%%%%%%%%%%%%%%%%%%%%%%%%%%%%%
\section{Summary and Outlook}\label{sc:sums}
%%%%%%%%%%%%%%%%%%%%%%%%%%%%%%%%%%%%%%%%%%%%%%%%%%%%%%%%%%%%%%%%%%%%%%%%%%%%%%%%%%%%%%%%%%%%%%%%%%%%%%%%%%%%

The origin of dark matter remains one of the most compelling mysteries in our understanding of the Universe today and the Large Hadron Collider is playing a central role in constraining some of its parameters. A suite of analyses looking for mono-$X$ plus missing transverse energy has already extended the exclusion bounds set by direct detection experiments. A deviation from SM in inclusive signatures like missing energy plus jets (plus leptons) may hint a discovery and, although these scheme has been developed with supersymmetry in mind, it has already been applied to other beyond-standard-model scenarios such as universal extra dimension models.

If LHC should discover general WIMP dark matter, it will be non-trivial to prove that it has the right properties. Future $e^+e^-$ colliders (ILC, CLIC) are expected to extend the LHC discovery potential and improve the identification of the underlying DM model. By providing more precise determination of model parameters, they will consequently shed light on the relic density, the direct detection rate and the WIMP annihilation processes. The complementarity between LHC and cosmo/astroparticle experiments lies in the uncorrelated systematics and the measurement of different model parameters. In the following years we expect a continuous interplay between particle physics experiments and astrophysical/cosmological observations.

%%%%%%%%%%%%%%%%%%%%%%%%%%%%%%%%%%%%%%%%%%%%%%%%%%%%%%%%%%%%%%%%%%%%%%%%%%%%%%%%%%%%%%%%%%%%%%%%%%%%%%%%%%%%
\section*{Acknowledgments}

The author acknowledges support by the Spanish Ministry of Economy and Competitiveness (MINECO) under the projects FPA2009-13234-C04-01 and FPA2012-39055-C02-01, by the Generalitat Valenciana through the project PROMETEO~II/2013-017 and by the Spanish National Research Council (CSIC) under the JAE-Doc program co-funded by the European Social Fund (ESF). 

%%%%%%%%%%%%%%%%%%%%%%%%%%%%%%%%%%%%%%%%%%%%%%%%%%%%%%%%%%%%%%%%%%%%%%%%%%%%%%%%%%%%%%%%%%%%%%%%%%%%%%%%%%%%
%%%%%%%%%%%%%%%%%%%%%%%%%%%%%%%%%%%%%%%%%%%%%%%%%%%%%%%%%%%%%%%%%%%%%%%%%%%%%%%%%%%%%%%%%%%%%%%%%%%%%%%%%%%%

\end{document}